\def\arxiv{}
\newif\iflinenums\linenumstrue

\def\prl{}

\def\print{preprint}
\newcommand{\citesupp}{\cite{ArXivSupplement}}
\newcommand{\citetsupp}{the Supplemental Material \cite{ArXivSupplement}}

\ifdefined\prl
  \linenumsfalse
  \renewcommand{\citesupp}{\cite{PRLSupplement}}
  \renewcommand{\citetsupp}{Supplemental Material \cite{PRLSupplement}}
  \ifdefined\supplemental
    \def\print{onecolumn}
  \else
    \def\print{preprint}
  \fi
\fi

\ifdefined\arxiv
  \linenumsfalse
  \renewcommand{\citesupp}{\cite{ArXivSupplement}}
  \renewcommand{\citetsupp}{Supplemental Material \cite{ArXivSupplement}}
  \ifdefined\supplemental
    \def\print{onecolumn}
  \else
    \def\print{reprint}
  \fi
\fi

\documentclass[aps,prl,\print,superscriptaddress,floatfix,nofootinbib,showkeys,showpacs]{revtex4-1}

\usepackage{graphicx}
\usepackage{aps_macros}
\usepackage[dvipsnames,svgnames]{xcolor} 
\usepackage{natbib} 
\usepackage{hyperref}
\usepackage{amsmath}
\usepackage{multirow}
\usepackage{accents}
\hypersetup{    
  colorlinks      = true,
  linkcolor       = {black},
  linkbordercolor = {white},
  citecolor       = {black},
  citebordercolor = {white},
  urlcolor        = {blue},
  urlbordercolor  = {white},
}
\usepackage[all]{hypcap}

\usepackage{soul} 
\usepackage{amsmath}
\usepackage{xspace}
\usepackage{lineno}

\graphicspath{{./}{../figs/}}

\newcommand{\ie}{i.e.\xspace}
\newcommand{\eg}{e.g.\xspace}

\newcommand{\New}[1]{#1}

\mathchardef\mhyphen="2D
\newcommand{\vect}[1]{\boldsymbol{#1}}
\newcommand{\roughly}{\ensuremath{ {\sim}\,} }

\newlength{\dhatheight}
\newcommand{\doublehat}[1]{%
    \settoheight{\dhatheight}{\ensuremath{\hat{#1}}}%
    \addtolength{\dhatheight}{-0.35ex}%
    \hat{\vphantom{\rule{1pt}{\dhatheight}}%
    \smash{\hat{#1}}}}

\newcommand{\unit}[1]{\ensuremath{\mathrm{\,#1}}\xspace}
\newcommand{\MeV}{\unit{MeV}}
\newcommand{\GeV}{\unit{GeV}}
\newcommand{\TeV}{\unit{TeV}}
\newcommand{\degree}{\ensuremath{{}^{\circ}}\xspace}

\newcommand{\cm}{\unit{cm}}

\newcommand{\second}{\unit{s}}

\newcommand{\photon}{\unit{ph}}
\newcommand{\sr}{\unit{sr}}

\newcommand{\cmcubes}{\ensuremath{\cm^{3}\second^{-1}}\xspace}
\newcommand{\GeVcmcube}{\ensuremath{\GeV\cm^{-3}}\xspace}

\newcommand{\tabref}[1]{Table~\ref{tab:#1}}

\newcommand{\figref}[1]{Figure~\ref{fig:#1}}

\newcommand{\eqnref}[1]{Equation~\eqref{eqn:#1}}

\newcommand{\Fermi}{\textit{Fermi}\xspace}

\newcommand{\aeff}{\ensuremath{A_{\rm eff}}}
\newcommand{\TS}{\ensuremath{\mathrm{TS}}\xspace}

\newcommand{\passseven}{\code{Pass\,7}}
\newcommand{\passsevenrep}{\code{Pass\,7 Reprocessed}}

\newcommand{\passeight}{\code{Pass\,8}}

\newcommand{\Prob}{\ensuremath{\mathcal{P}}\xspace}

\newcommand{\ProbJi}{\ensuremath{\mathcal{P}(J_{i})}\xspace}
\newcommand{\like}{\ensuremath{\mathcal{L}}\xspace} 
\newcommand{\pseudolike}{ {\tilde{\like}} \xspace}   
\newcommand{\given}{\ensuremath{ \,|\, }\xspace}

\newcommand{\data}{ \ensuremath{ \mathcal{D} }\xspace } 

\newcommand{\params}{\ensuremath{\vect{\alpha}}\xspace}
\newcommand{\sig}{\ensuremath{\mu}\xspace}
\newcommand{\bkg}{\ensuremath{\theta}\xspace}
\newcommand{\interest}{\ensuremath{\vect{\sig}}\xspace}
\newcommand{\nuisance}{\ensuremath{\vect{\bkg}}\xspace}

\newcommand{\Jlike}{\ensuremath{\like_{J}}\xspace}
\newcommand{\Jsigma}{\ensuremath{\sigma_{i}}\xspace}
\newcommand{\Jtrue}{\Ji}
\newcommand{\Jobs}{\ensuremath{J_{\rm{obs}}}\xspace}
\newcommand{\Jobsi}{\ensuremath{J_{\rm{obs},i}}\xspace}
\newcommand{\logtenJtrue}{\logtenJi}
\newcommand{\logtenJobs}{\ensuremath{ {\log_{10}{(\Jobsi)}} }\xspace}

\newcommand{\Ji}{\ensuremath{J_i}\xspace}

\newcommand{\logtenJi}{\ensuremath{{\log_{10}{(J_i)}}}\xspace}

\newcommand{\CL}{CL\xspace}

\newcommand{\code}[1]{\texttt{#1}\xspace}

\newcommand{\DMFIT}{\code{DMFIT}}
\newcommand{\Pythia}{\code{Pythia}}

\newcommand{\DM}{\ensuremath{\mathrm{DM}}}
\newcommand{\mDM}{\ensuremath{m_\DM}\xspace}

\newcommand{\sigmav}{\ensuremath{\langle \sigma v \rangle}\xspace}

\newcommand{\uubar}{\ensuremath{u \bar u}\xspace}

\newcommand{\bbbar}{\ensuremath{b \bar b}\xspace}

\newcommand{\ww}{\ensuremath{W^{+}W^{-}}\xspace}

\newcommand{\ee}{\ensuremath{e^{+}e^{-}}\xspace}
\newcommand{\mumu}{\ensuremath{\mu^{+}\mu^{-}}\xspace}
\newcommand{\tautau}{\ensuremath{\tau^{+}\tau^{-}}\xspace}

\newcommand{\relic}{\ensuremath{2.2\times10^{-26}\cm^{3}\second^{-1}}\xspace}

\newcommand{\Jfactor}{J-factor\xspace}
\newcommand{\Jfactors}{J-factors\xspace}

\newcommand{\NRAND}{300\xspace}

\providecommand\physrep{\ref@jnl{Phys.~Rep.}}%
\providecommand\apjs{\ref@jnl{ApJS}}%
\providecommand{\jcap}{\ref@jnl{JCAP}}%

\iflinenums
  \linenumbers
\fi

\begin{document}

\title{Searching for Dark Matter Annihilation from Milky Way Dwarf
  Spheroidal Galaxies with Six Years of Fermi-LAT Data} 

\date{\today}
\author{M.~Ackermann}
\affiliation{Deutsches Elektronen Synchrotron DESY, D-15738 Zeuthen, Germany}
\author{A.~Albert}
\affiliation{W. W. Hansen Experimental Physics Laboratory, Kavli Institute for Particle Astrophysics and Cosmology, Department of Physics and SLAC National Accelerator Laboratory, Stanford University, Stanford, CA 94305, USA}
\author{B.~Anderson}
\email{brandon.anderson@fysik.su.se}
\affiliation{Department of Physics, Stockholm University, AlbaNova, SE-106 91 Stockholm, Sweden}
\affiliation{The Oskar Klein Centre for Cosmoparticle Physics, AlbaNova, SE-106 91 Stockholm, Sweden}
\author{W.~B.~Atwood}
\affiliation{Santa Cruz Institute for Particle Physics, Department of Physics and Department of Astronomy and Astrophysics, University of California at Santa Cruz, Santa Cruz, CA 95064, USA}
\author{L.~Baldini}
\affiliation{Universit\`a di Pisa and Istituto Nazionale di Fisica Nucleare, Sezione di Pisa I-56127 Pisa, Italy}
\affiliation{W. W. Hansen Experimental Physics Laboratory, Kavli Institute for Particle Astrophysics and Cosmology, Department of Physics and SLAC National Accelerator Laboratory, Stanford University, Stanford, CA 94305, USA}
\author{G.~Barbiellini}
\affiliation{Istituto Nazionale di Fisica Nucleare, Sezione di Trieste, I-34127 Trieste, Italy}
\affiliation{Dipartimento di Fisica, Universit\`a di Trieste, I-34127 Trieste, Italy}
\author{D.~Bastieri}
\affiliation{Istituto Nazionale di Fisica Nucleare, Sezione di Padova, I-35131 Padova, Italy}
\affiliation{Dipartimento di Fisica e Astronomia ``G. Galilei'', Universit\`a di Padova, I-35131 Padova, Italy}
\author{K.~Bechtol}
\affiliation{Dept.  of  Physics  and  Wisconsin  IceCube  Particle  Astrophysics  Center, University  of  Wisconsin, Madison,  WI  53706, USA}
\author{R.~Bellazzini}
\affiliation{Istituto Nazionale di Fisica Nucleare, Sezione di Pisa, I-56127 Pisa, Italy}
\author{E.~Bissaldi}
\affiliation{Istituto Nazionale di Fisica Nucleare, Sezione di Bari, I-70126 Bari, Italy}
% ...
\collaboration{The Fermi-LAT Collaboration}
\noaffiliation
\begin{abstract}
  The dwarf spheroidal satellite galaxies (dSphs) of the Milky Way are
  some of the most dark matter (DM) dominated objects known.  We
  report on gamma-ray observations of Milky Way dSphs based on 6 years
  of \Fermi Large Area Telescope data processed with the new
  \passeight event-level analysis.  None of the dSphs are
  significantly detected in gamma rays, and we present upper limits on
  the DM annihilation cross section from a combined analysis of 15
  dSphs.  These constraints are among the strongest and most robust to
  date and lie below the canonical thermal relic cross section for DM
  of mass $\lesssim 100\GeV$ annihilating via quark and $\tau$-lepton
  channels.
\end{abstract}

\maketitle 

\section{Introduction}
\label{sec:intro}

Approximately 26\% of the energy density of the universe is composed
of non-baryonic cold dark matter (DM)~\cite{Adam:2015rua}.  Weakly
interacting massive particles (WIMPs) are an attractive candidate to
constitute some or all of
DM~\cite{Jungman:1995df,Bergstrom:2000pn,Bertone:2004pz}.
\New{The relic abundance of WIMPs is determined by their
  annihilation cross section at freeze-out \cite{Steigman:2012nb}, and
  the characteristic weak-scale cross sections of WIMPs can naturally
  produce a relic abundance equal to the observed abundance of DM.}
Self-annihilation of WIMPs would continue today in regions of high DM
density and result in the production of energetic Standard Model
particles.  The large mass of the WIMP (\mDM) permits the production
of gamma rays observable by the \Fermi Large Area Telescope (LAT),
which is sensitive to energies ranging from 20 \MeV to ${>}\,300\GeV$.

Kinematic data indicate that the dwarf spheroidal satellite galaxies
(dSphs) of the Milky Way contain a substantial DM
component~\cite{Mateo:1998wg,McConnachie:2012vd}.  The gamma-ray
signal flux at the LAT, $\phi_s$ ($\photon \cm^{-2} \second^{-1}$),
expected from the annihilation of DM with a density distribution
$\rho_{\DM}(\vect{r})$ is given by
\begin{equation}
\begin{aligned}
       \phi_s(\Delta\Omega) =
    & \underbrace{ \frac{1}{4\pi} \frac{\sigmav}{2m_{\DM}^{2}}\int^{E_{\max}}_{E_{\min}}\frac{\text{d}N_{\gamma}}{\text{d}E_{\gamma}}\text{d}E_{\gamma}}_{\rm particle~physics}\\
    &    \times
    \underbrace{\vphantom{\int_{E_{\min}}} \int_{\Delta\Omega}\int_{\rm l.o.s.}\rho_{\DM}^{2}(\vect{r})\text{d}l\text{d}\Omega '}_{\rm \Jfactor}\,.
\end{aligned}
\label{eqn:annihilation}
\end{equation}
Here, the first term is dependent on the particle physics properties --- \ie, the thermally-averaged annihilation cross section, \sigmav, the particle mass, \mDM, and the differential gamma-ray yield per annihilation, $\text{d}N_\gamma/\text{d}E_\gamma$, integrated over the experimental energy range.%
\footnote{Strictly speaking, the differential yield per annihilation in Equation (\ref{eqn:annihilation}) is a sum of differential yields into specific final states: ${\text{d}N_\gamma/\text{d}E_\gamma = \sum_f B_f\; \text{d}N^f_\gamma/\text{d}E_\gamma}$, where $B_f$ is the branching fraction into final state $f$. Here, we make use of Equation (\ref{eqn:annihilation}) in the context of single final states only.}
The second term, known as the \Jfactor, is the line-of-sight (l.o.s.) integral through the DM distribution integrated over a solid angle, $\Delta\Omega$. 

Milky Way dSphs can give rise to \Jfactors  in excess of $10^{19} \GeV^{2} \cm^{-5}$ \cite{Martinez:2013ioa,Geringer-Sameth:2014yza}, which, coupled with their lack of non-thermal astrophysical processes, makes them good targets for DM searches via gamma rays.  Gamma-ray searches for dSphs yield some of the most stringent constraints on \sigmav, particularly when multiple dSphs are analyzed together using a joint likelihood technique~\cite{Ackermann:2011wa,GeringerSameth:2011iw,Mazziotta:2012ux,Ackermann:2013yva,Geringer-Sameth:2014qqa,2015arXiv150203081A}.
Limits on \sigmav derived from observations of dSphs have begun to probe the low-\mDM parameter space for which the WIMP abundance matches the observed DM relic density.  

In contrast, DM searches in the Galactic center take advantage of a \Jfactor that is $\mathcal{O}(100)$ times larger, although gamma-ray emission from non-thermal processes makes a bright, structured background.  Several studies of the Galactic center interpret an excess of gamma rays with respect to modeled astrophysical backgrounds as a signal of 20 to 50\GeV WIMPs annihilating via the $\bbbar$ channel \cite{Daylan:2014rsa,Gordon:2013vta,Abazajian:2014fta,Calore:2014xka}.  
Coincidentally, the largest deviation from expected background in some
previous studies of dSphs occurred for a similar set of WIMP characteristics; however, this deviation was not statistically significant \cite{Ackermann:2013yva}.

Using a new LAT event-level analysis, known as \passeight, we
re-examine the sample of 25 Milky Way dSphs from
\citet{Ackermann:2013yva} using six years of LAT data.  The \passeight
data benefits from an improved point-spread function (PSF), effective
area, and energy reach.  More accurate Monte Carlo simulations of the
detector and the environment in low-Earth orbit have reduced the
systematic uncertainty in the LAT instrument response functions
(IRFs)~\cite{Atwood:2013rka}.  Within the standard photon classes,
\passeight offers \emph{event types}, subdivisions based on
event-by-event uncertainties in the directional and energy
measurements, which can increase the sensitivity of likelihood-based
analyses.
In this work we use a set of four PSF event-type selections that
subdivide the events in our data sample according to the quality of
their directional reconstruction.  In addition to the improvements
from \passeight, we employ the updated third LAT source catalog
(3FGL), based on four years of \passsevenrep data, to model
point-like background sources~\cite{Ackermann:2015hja}.  Together,
these improvements, along with an additional two years of data taking,
lead to a predicted increase in sensitivity of 70\% relative to the
four-year analysis of \citet{Ackermann:2013yva} for the \bbbar channel
at 100\GeV.  More details on \passeight and other aspects of this
analysis can be found in \citetsupp.
  
\section{LAT data selection}
We examine six years of LAT data (2008-08-04 to 2014-08-05) selecting
\passeight SOURCE-class events in the energy range between 500\MeV and
500\GeV.  We selected the 500\MeV lower limit to mitigate the impact
of leakage from the bright limb of the Earth because the PSF broadens
considerably below that energy.
To further avoid contamination from terrestrial gamma rays, events
with zenith angles larger than 100\degree are rejected.  We also
remove time intervals around bright GRBs and solar flares following
the prescription used for the 3FGL catalog.  We extract from this data
set $10\degree {\times} 10\degree$ square regions of interest (ROIs)
in Galactic coordinates centered at the position of each dSph
specified in \tabref{dsphs}.

At a given energy, \New{20\%--40\%} of the events classified as photons in
our six-year \passeight data set are shared with the analysis of
\citet{Ackermann:2013yva}.  The low fraction of shared events can be
attributed primarily to the larger time range used for the present analysis
(four versus six years) and the increase in gamma-ray acceptance of the
P8R2\_SOURCE event class relative to P7REP\_CLEAN.  \New{At most, the
\passseven events can represent 35\%--50\% of the new, larger sample.  
Migration of the individual reconstructed events, \New{particularly residual 
cosmic rays,} across ROI and class selection boundaries further reduces 
the overlap, making the two analyses nearly statistically independent \citesupp.}


\section{J-factors for dwarf spheroidal galaxies}

The DM content of dSphs can be determined through dynamical modeling
of their stellar density and velocity dispersion
profiles~\cite{Walker:2012td,Battaglia:2013wqa,Strigari:2012gn}.
Recent studies have shown that an accurate estimate of the dynamical
mass of a dSph can be derived from measurements of the average stellar
velocity dispersion and half-light radius
alone~\cite{Walker:2009zp,Wolf:2009tu}.  The total mass within the
half-light radius and the integrated \Jfactor have been found to be
fairly insensitive to the assumed DM density
profile~\cite{Martinez:2009jh,Strigari:2012gn,Ackermann:2013yva}.  We
assume that the DM distribution in dSphs follows a Navarro-Frenk-White
(NFW) profile~\cite{Navarro:1996gj},
\begin{equation}
 \label{eq:nfw}
 \rho_{\DM}(r) = \frac{\rho_0 r_s^3}{ r ( r_s + r)^2 },
\end{equation}
where $r_s$ and $\rho_0$ are the NFW scale radius and characteristic density, respectively.
We take J-factors and other physical properties for the Milky Way
dSphs from \citet{Ackermann:2013yva} (and references therein).

\section{Data analysis}

We perform a binned Poisson maximum-likelihood analysis in 24 bins of
energy,\footnote{Constraints are insensitive to finer binning.}
logarithmically spaced from 500\MeV to 500\GeV, and an 0.1\degree
angular pixelization.  \New{The low-energy bound of 500\MeV is
  selected to mitigate the impact of leakage from the bright limb of
  the Earth because the PSF broadens considerably below that
  energy. The high-energy bound of 500\GeV is chosen to mitigate the
  effect of the increasing residual charged-particle background at
  higher energies~\citep{Ackermann:2014usa}.}  The data were analyzed
with the Fermi Science
Tools\footnote{\url{http://fermi.gsfc.nasa.gov/ssc/data/analysis/software}}
version 10-01-01
and the P8R2\_SOURCE\_V6 IRFs.
Our diffuse background model includes a structured Galactic component
and a spatially isotropic component that represents both extragalactic
emission and residual particle contamination.%
\footnote{\url{http://fermi.gsfc.nasa.gov/ssc/data/access/lat/BackgroundModels.html}}
Because the energy resolution of the LAT was not accounted for when
fitting the Galactic diffuse model, differences in response (energy
resolution and effective area) between IRF sets lead to different
measured intensities for this component.  Thus, a small
energy-dependent scaling has been applied to the \passsevenrep
Galactic diffuse model.  \New{Changes with respect to the
  \passsevenrep model are less than 5\% above 100\MeV.  Details on the
  derivation of the rescaled model are given in \citesupp.} The
gamma-ray characteristics of nearby point-like sources are taken from
the 3FGL catalog~\cite{Ackermann:2015hja}.

We perform a bin-by-bin likelihood analysis of the gamma-ray emission
coincident with each dSph following the procedure of
\citet{Ackermann:2013yva}.  
\New{The flux normalizations of the Galactic diffuse and isotropic components and 3FGL catalog sources within the $10^{\circ}\times 10^{\circ}$ ROI were fit simultaneously in a binned likelihood analysis over the broadband energy range from 500\MeV to 500\GeV.}
\New{The normalizations of the background sources are insensitive to the inclusion of a putative power-law source at the locations of the dSphs, which is consistent with the lack of any strong signal associated with the dSphs.}
Fixing the normalizations of the background sources \New{with} the
broad-band fit \New{before fitting each bin individually} avoids
numerical instability resulting from the fine binning in energy and
the degeneracy of the diffuse background components at high Galactic
latitudes.

\New{After fixing the background normalizations,} we scan the
likelihood as a function of the flux normalization of the putative DM
signal independently in each energy bin (this procedure is similar to
that used to evaluate the spectral energy distribution of a
source).  \New{Within each bin, we model the putative dSph source with
a power-law spectral model (dN/d$E$ $\propto E^{-\Gamma}$) with
spectral index of $\Gamma = 2$.}  By analyzing each energy bin
separately, we avoid selecting a single spectral shape to span the
entire energy range at the expense of introducing additional degrees
of freedom into the fit.

While the bin-by-bin likelihood function is essentially independent of
spectral assumptions, it does depend on the spatial model of the DM
distribution in the dSphs.  We model the dSphs with spatially extended
NFW DM density profiles projected along the line of sight.  The
angular extent of the emission profile for each dSph is set by the
scale radius of its DM halo, which contains approximately 90\% of
the total annihilation flux.  We use the set of DM halo scale radii from
\citet{Ackermann:2013yva}, which span a range of subtended angles
between 0.1\degree and 0.4\degree.


We test a wide range of DM annihilation hypotheses by using predicted gamma-ray spectra to tie the signal normalization across the energy bins.  Spectra for DM annihilation are generated with the \DMFIT package based on
\Pythia~8.165~\cite{Jeltema:2008hf,Sjostrand:2007gs,Ackermann:2013yva}.
We reconstruct a broad-band likelihood function by multiplying the
bin-by-bin likelihood functions
evaluated at the predicted fluxes for a given DM model.

We combine the broad-band likelihood functions across 15 of the observed
dSphs\footnote{Selected to have kinematically determined \Jfactors and
  avoid ROI overlap.  The set is identical to that in
  \citet{Ackermann:2013yva}.}  and include statistical uncertainties on
the \Jfactors of each dSph by adding an additional \Jfactor likelihood
term to the binned Poisson likelihood for the LAT data. The \Jfactor
likelihood for target $i$ is given by
\begin{equation}\label{eqn:jfactor_likelihood}
\begin{aligned}
  \Jlike(\Jtrue \given \Jobsi, \Jsigma) &= \frac{1}{\ln(10) \Jobsi
    \sqrt{2 \pi} \sigma_i} \\
  &\times e^{-\left(\logtenJtrue-\logtenJobs\right)^2/2\sigma_i^2},
\end{aligned}
\end{equation}
where \Jtrue is the true value of the \Jfactor and \Jobsi is the
measured \Jfactor with error \Jsigma.  This parameterization of the
\Jfactor likelihood is obtained by fitting a log-normal function with
peak value \Jobsi to the posterior distribution for each \Jfactor as
derived by \citet{Martinez:2013ioa}, providing a reasonable way to
quantify the uncertainties on the \Jfactors.
This approach is a slight modification of the approach in
\citet{Ackermann:2011wa,Ackermann:2013yva}, where an effective
likelihood was derived considering a flat prior on the \Jfactors.  We
note that the \Jfactor correction is only intended to incorporate the
\emph{statistical} uncertainty in the \Jfactors, and not the
systematic uncertainty resulting from the fitting procedure or choice
of priors \citesupp.  More details on the derivation of the \Jfactor
likelihood and the effects of systematic uncertainties can be found
in~\citetsupp.

Combining the broad-band gamma-ray and \Jfactor likelihood functions, our likelihood function for target $i$ becomes,
\begin{equation}\label{eqn:full_likelihood}
\begin{aligned}
\pseudolike_i(\interest,\nuisance_i =\{\params_{i},J_{i}\}\given
\data_i) = &\like_i(\interest,\nuisance_i \given \data_i) 
\Jlike(\Jtrue \given \Jobsi, \Jsigma).
\end{aligned}
\end{equation}
Here, $\interest$ are the parameters of the DM model, $\nuisance_{i}$
is the set of nuisance parameters that includes both nuisance
parameters from the LAT analysis ($\params_{i}$) and the dSph \Jfactor
(\Jtrue), and $\data_{i}$ is the gamma-ray data.  We incorporate
additional information about the event-wise quality of the angular
reconstruction by forming the LAT likelihood function ($\like_i$) from
the product of likelihood functions for four PSF event types.  The four PSF
event types (PSF0, PSF1, PSF2, and PSF3) subdivide the events in the
SOURCE-class data set into exclusive partitions ($\data_{i,j}$) in
order of decreasing uncertainty on the direction measurement.  The
resulting joint LAT likelihood function is given by
\begin{equation}\label{eqn:psf_likelihood}
\begin{aligned}
  \like_i(\interest,\nuisance_i \given \data_{i}) = &
  \prod_{j}\like_i(\interest,\nuisance_i \given \data_{i,j}).
\end{aligned}
\end{equation}
The spectral and spatial model of gamma-ray counts for each event type
partition is evaluated using a set of IRFs computed for that class and
type selection.

We evaluate the significance of DM hypotheses using a test statistic
(\TS) defined as
\begin{equation}
  \TS = -2~\rm ln\left(\frac{\like(\interest_{0},\hat{\nuisance}
      \given\data)}
    {\like(\hat{\interest},\hat{\nuisance}\given\data)}\right),
\label{eqn:TS}
\end{equation}
where $\interest_{0}$ are the parameters of the null (no DM)
hypothesis and $\hat{\interest}$ \New{and $\hat{\nuisance}$} are the
best-fit parameters under the DM hypothesis.  \like can here be either
the likelihood for an individual dSph or the joint likelihood for the
dSphs in our combined sample.  \New{We note that following the
  methodology of \citet{Ackermann:2013yva} we use background
  parameters ($\hat{\nuisance}$) derived under the hypothesis of a DM
  source with a $\Gamma=2$ power-law spectrum when evaluating both the
  null and DM hypotheses.  This is a good approximation as long as the
  best-fit signal is small relative to the background in the ROI.}
\noindent
Based on the asymptotic theorem of~\citet{Chernoff:1954}, the \TS can
be converted to a significance based on a mixture of $\chi^2$
distributions.  The validity of this assumption is examined further in
\citetsupp.

\section{Results}

\begin{figure*}[ht]
  \centering
  \includegraphics[width=0.49\textwidth]{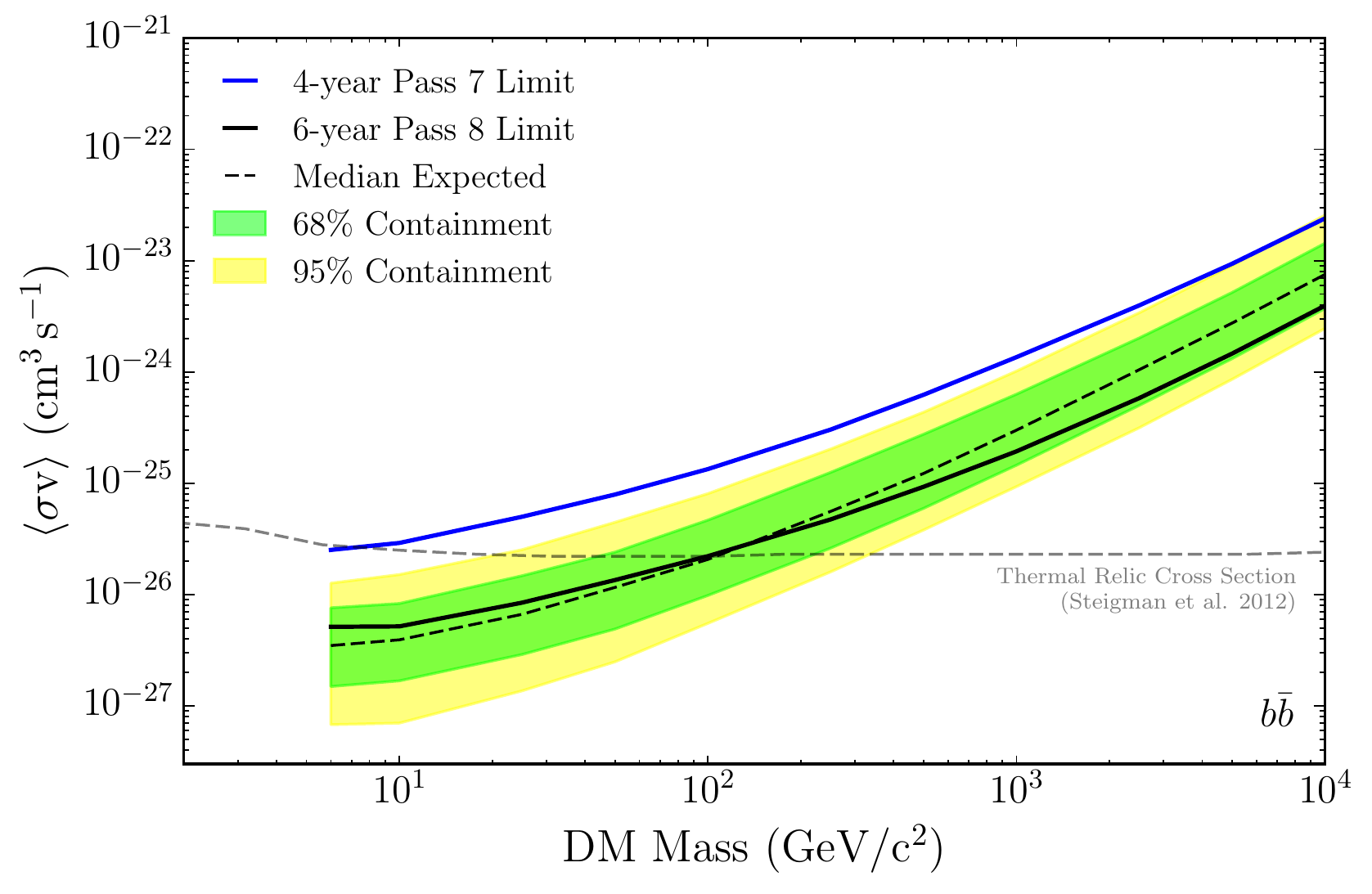}
  \includegraphics[width=0.49\textwidth]{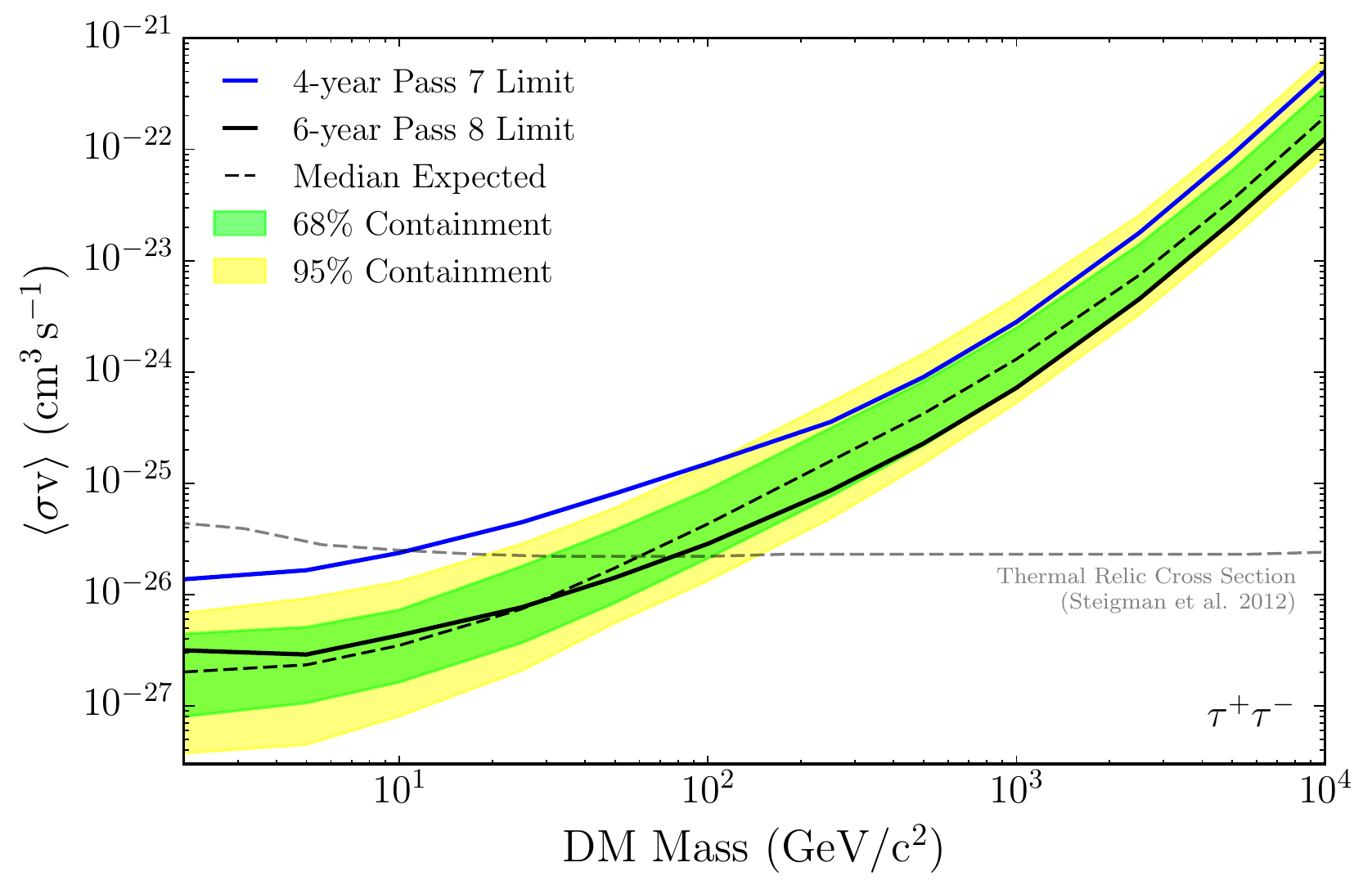}
  \caption{Constraints on the DM annihilation cross section at 95\%
    \CL for the \bbbar (left) and \tautau (right) channels derived
    from a combined analysis of 15 dSphs.  Bands for the expected
    sensitivity are calculated by repeating the same analysis on
    \NRAND randomly selected sets of high-Galactic-latitude blank
    fields in the LAT data.  The dashed line shows the median expected
    sensitivity while the bands represent the 68\% and 95\% quantiles.
    For each set of random locations, nominal \Jfactors are randomized
    in accord with their measurement uncertainties.  The solid blue
    curve shows the limits derived from a previous analysis of
    four years of \passsevenrep data and the same sample of
    15 dSphs \cite{Ackermann:2013yva}.  The dashed gray curve in this
    and subsequent figures corresponds to the thermal relic cross
    section from \citet{Steigman:2012nb}.}\label{fig:fig1}
\end{figure*}

We find no significant gamma-ray excess associated with the Milky Way
dSphs when analyzed individually or as a population.  In the combined
analysis of 15 dSphs, the largest deviation from the background-only
hypothesis has $\TS = 1.3$ occurring for $\mDM = 2 \GeV$ annihilating
through the \ee channel.  Among the dSphs in our combined analysis,
the dSph with the largest individual significance is Sculptor with
$\TS = 4.3$ for \mDM = 5\GeV annihilating through the \mumu channel.
The maximum \TS of our combined analysis is well below the threshold
set for gamma-ray source detection and is completely consistent with a
background fluctuation~\cite{Ackermann:2015hja}.  We set upper limits
on \sigmav at 95\% confidence level (\CL) for WIMPs with \mDM between
2 \GeV and 10 \TeV annihilating into six different standard model
channels (\bbbar, \tautau, \mumu, \ee, \ww, \uubar).
\footnote{Results for all channels as well as bin-by-bin likelihood
  functions for each target are available in machine-readable format
  at: \url{http://www-glast.stanford.edu/pub_data/1048/}.}
\figref{fig1} shows the comparison of the limits for the \bbbar and
\tautau channels with expectation bands derived from the analysis of
300 randomly selected sets of blank fields.  Sets of blank fields are
generated by choosing random sky positions with $|b| > 30\degree$ that
are centered at least $0.5\degree$ from 3FGL catalog sources.  We
additionally require fields within each set to be separated by at
least $7\degree$.  Our expected limit bands are evaluated with the
3FGL source catalog based on four years of \passsevenrep data and
account for the influence of new sources present in the six-year
\passeight data set.

Comparing with the results of \citet{Ackermann:2013yva}, we find a
factor of 3--5 improvement in the limits for all channels using six
years of \passeight data and the same sample of 15 dSphs.  The larger
data set as well as the gains in the LAT instrument performance
enabled by \passeight both contribute to the increased sensitivity of
the present analysis.  An additional 30--40\% improvement in the limit
can be attributed to the modified functional form chosen for the
\Jfactor likelihood (Equation \ref{eqn:jfactor_likelihood}).
Statistical fluctuations in the \passeight data set also play a
substantial role.  Because the \passeight six-year and \passsevenrep
four-year event samples have a shared fraction of only \New{20--40\%}, the
two analyses are nearly statistically independent.  For masses below
100\GeV, the upper limits of \citet{Ackermann:2013yva} were near the
95\% upper bound of the expected sensitivity band while the limits in
the present analysis are within one standard deviation
of the median expectation value.

Uncertainties in the LAT IRFs, modeling of the diffuse background, and estimation of \Jfactors all contribute systematic errors to this analysis.  By
examining maximal variations of each contributor, we find that at 100 \GeV they lead to ${\pm}9\%$, ${\pm}8\%$, and ${\pm}33\%$ shifts in our
limits, respectively (see \citetsupp).

\begin{figure*}[ht]
  \centering
  \includegraphics[width=0.49\textwidth]{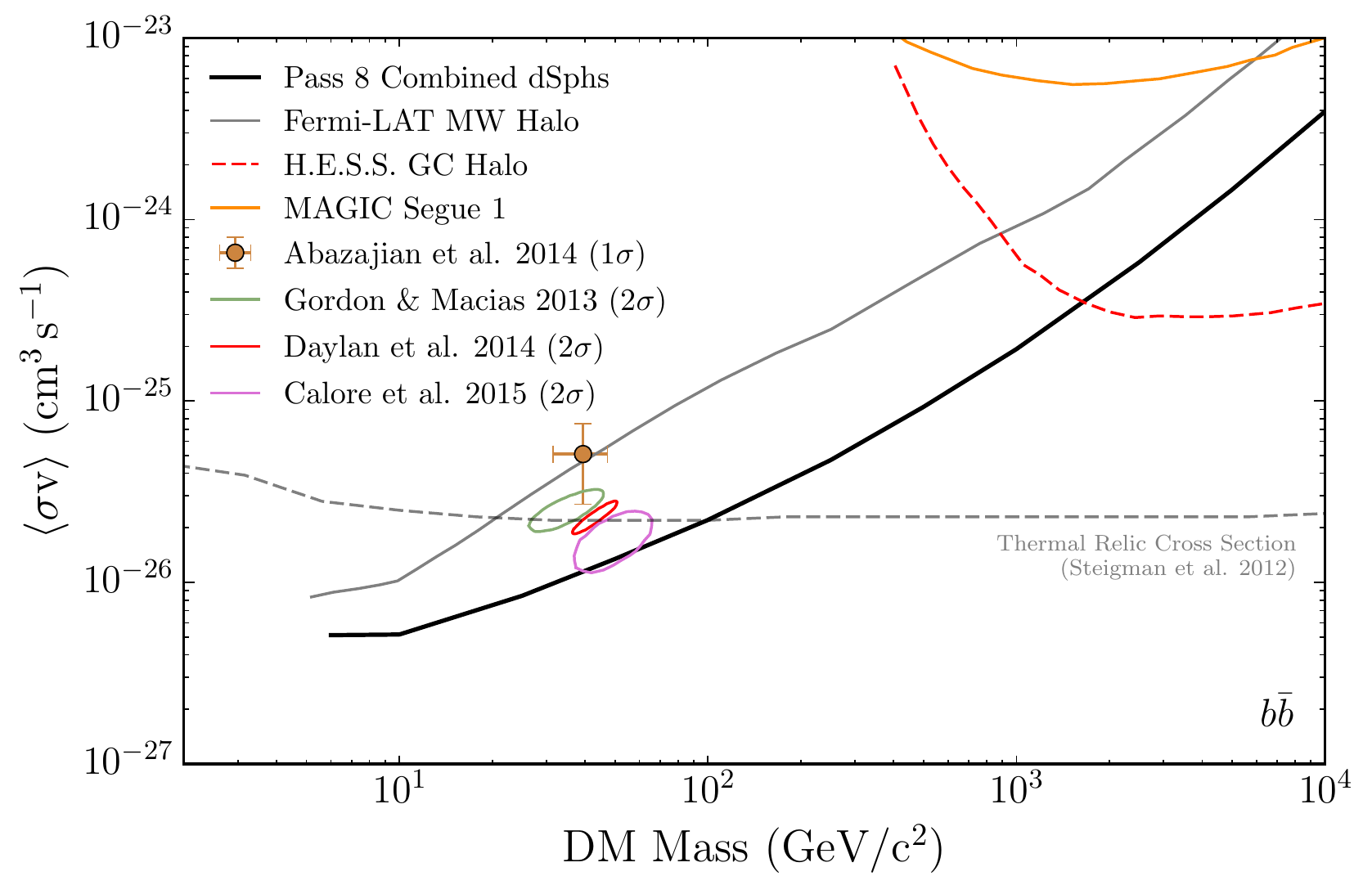}
  \includegraphics[width=0.49\textwidth]{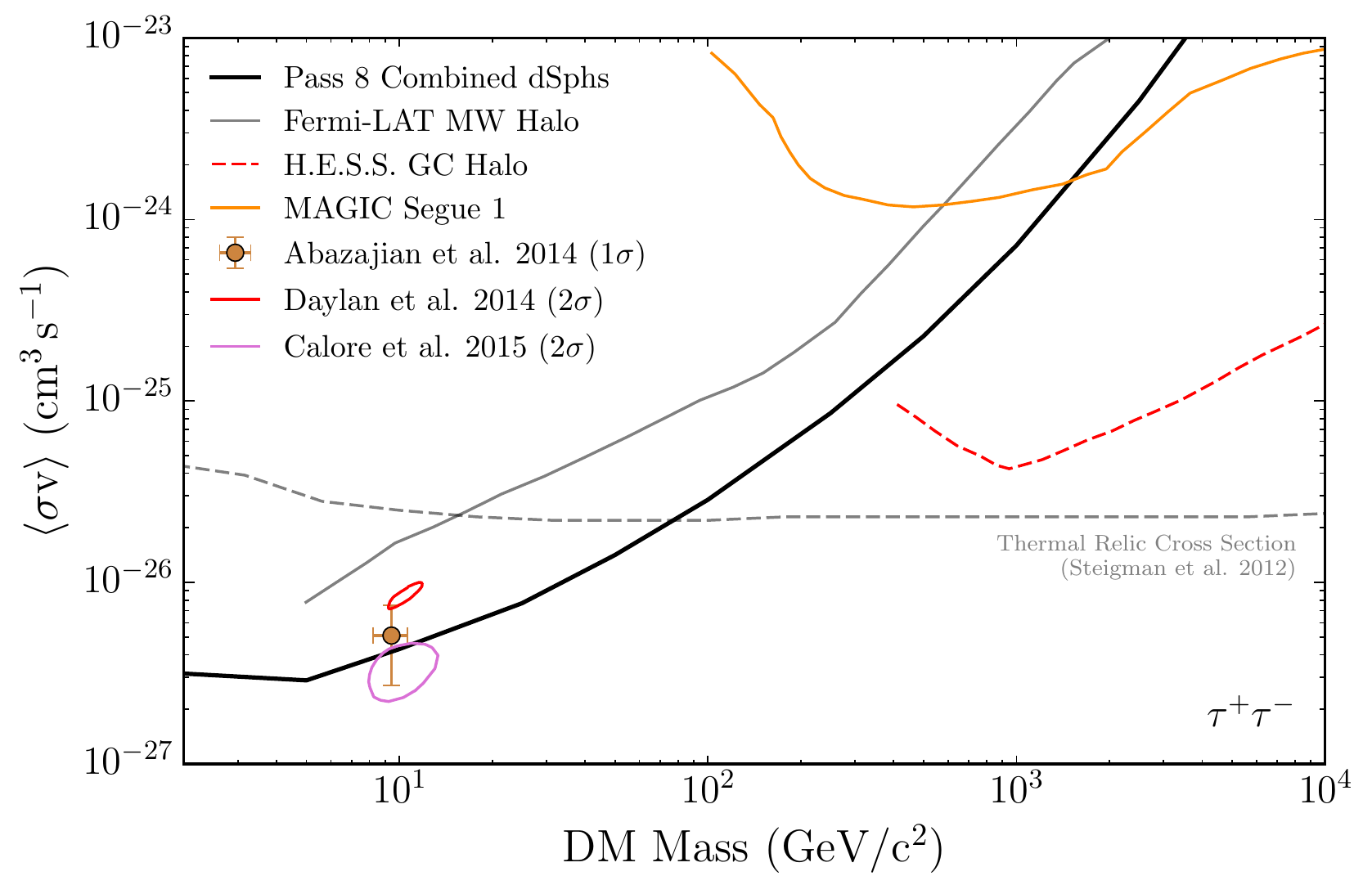}
  \caption{Comparison of constraints on the DM annihilation cross
    section for the \bbbar (left) and \tautau (right) channels from
    this work with previously published constraints from LAT analysis
    of the Milky Way halo ($3\sigma$ limit) \cite{Ackermann:2012rg},
    112 hours of observations of the Galactic Center with H.E.S.S.
    \cite{Abramowski:2011hc}, and 157.9 hours of observations of
    {Segue~1} with MAGIC \cite{Aleksic:2013xea}.  \New{Pure
      annihilation channel limits for the Galactic Center
      H.E.S.S. observations are taken from \citet{Abazajian:2011ak}
      and assume an Einasto Milky Way density profile with $\rho_\odot
      = 0.389$\GeVcmcube.}  Closed contours and the marker with error
    bars show the best-fit cross section and mass from several
    interpretations of the Galactic center excess
    \cite{Abazajian:2014fta,Calore:2014xka,Gordon:2013vta,Daylan:2014rsa}.}\label{fig:fig2}
\end{figure*}
 
Our results begin to constrain some of the preferred parameter space
for a DM interpretation of a gamma-ray excess in the Galactic center
region~\cite{Gordon:2013vta,Abazajian:2014fta,Daylan:2014rsa,Calore:2014xka}.
As shown in \figref{fig2}, for interpretations assuming a \bbbar final
state, the best-fit models lie in a region of parameter space slightly
above the 95\% \CL upper limit from this analysis, with an
annihilation cross section in the range of (1--3)$\times
10^{-26}\cmcubes$ and \mDM between 25 and 50\GeV.  However,
uncertainties in the structure of the Galactic DM distribution can
significantly enlarge the best-fit regions of \sigmav, channel, and \mDM \citep{2014arXiv1411.2592A}.

In conclusion, we present a combined analysis of 15 Milky Way dSphs
using a new and improved LAT data set processed with the \passeight
event-level analysis.  We exclude the thermal relic annihilation cross
section ($\roughly \relic$) for WIMPs with ${\mDM \lesssim 100\GeV}$
annihilating through the quark and $\tau$-lepton channels.  Our
results also constrain DM particles with \mDM above 100\GeV surpassing
the best limits from Imaging Atmospheric Cherenkov Telescopes for
masses up to \New{$\roughly 1\TeV$ for quark channels and $\roughly
  300\GeV$ for the $\tau$-lepton channel.}  These constraints include
the statistical uncertainty on the DM content of the dSphs.  The
future sensitivity to DM annihilation in dSphs will benefit from
additional LAT data taking and the discovery of new dSphs with
upcoming optical surveys such as the Dark Energy Survey
\cite{Abbott:2005bi} and the Large Synoptic Survey Telescope
\cite{Ivezic:2008fe}.

\begin{table}[h] \scriptsize
\caption{ \label{tab:dsphs} Properties of Milky Way dSphs.}
\begin{ruledtabular}
\begin{tabular}{ l r r c c l }
  Name                      & $\ell$\footnotemark[1] & $b$\footnotemark[1] & Distance & $\log_{10}({\Jobs})$\footnotemark[2] &  Ref. \\
  & ($\deg$) & ($\deg$) & (kpc) & ($\log_{10}[\GeV^2 \cm^{-5}]$) &   \\
  \hline
  Bootes I                  & 358.1  & 69.6   & 66     & $18.8 \pm 0.22$ & \cite{Dall'Ora:2006pt} \\
  Canes Venatici II         & 113.6  & 82.7   & 160    & $17.9 \pm 0.25$ & \cite{Simon:2007dq} \\
  Carina                    & 260.1  & $-$22.2  & 105    & $18.1 \pm 0.23$ & \cite{Walker:2008ax} \\
  Coma Berenices            & 241.9  & 83.6   & 44     & $19.0 \pm 0.25$ & \cite{Simon:2007dq} \\
  Draco                     & 86.4   & 34.7   & 76     & $18.8 \pm 0.16$ & \cite{Munoz:2005be} \\
  Fornax                    & 237.1  & $-$65.7  & 147    & $18.2 \pm 0.21$ & \cite{Walker:2008ax} \\
  Hercules                  & 28.7   & 36.9   & 132    & $18.1 \pm 0.25$ & \cite{Simon:2007dq} \\
  Leo II                    & 220.2  & 67.2   & 233    & $17.6 \pm 0.18$ & \cite{Koch:2007ye} \\
  Leo IV                    & 265.4  & 56.5   & 154    & $17.9 \pm 0.28$ & \cite{Simon:2007dq} \\
  Sculptor                  & 287.5  & $-$83.2  & 86     & $18.6 \pm 0.18$ & \cite{Walker:2008ax} \\
  Segue 1                   & 220.5  & 50.4   & 23     & $19.5 \pm 0.29$ & \cite{Simon:2010ek} \\
  Sextans                   & 243.5  & 42.3   & 86     & $18.4 \pm 0.27$ & \cite{Walker:2008ax} \\
  Ursa Major II             & 152.5  & 37.4   & 32     & $19.3 \pm 0.28$ & \cite{Simon:2007dq} \\
  Ursa Minor                & 105.0  & 44.8   & 76     & $18.8 \pm 0.19$ & \cite{Munoz:2005be} \\
  Willman 1                 & 158.6  & 56.8   & 38     & $19.1 \pm 0.31$ & \cite{Willman:2010gy} \\
\hline
  Bootes II \footnotemark[3]  & 353.7  & 68.9   & 42    & -- & -- \\
  Bootes III                            & 35.4    & 75.4   & 47    & -- & -- \\
  Canes Venatici I                 & 74.3    & 79.8   & 218  & $17.7 \pm 0.26$ &  \cite{Simon:2007dq} \\
  Canis Major                        & 240.0  & $-$8.0    & 7      & -- & -- \\
  Leo I                                   & 226.0  & 49.1   & 254   & $17.7 \pm 0.18$ & \cite{Mateo:2007xh} \\
  Leo V                                 & 261.9  & 58.5   & 178   & -- & -- \\
  Pisces II                             & 79.2    & $-$47.1  & 182   & -- & -- \\
  Sagittarius                         & 5.6      & $-$14.2  & 26     & -- & -- \\
  Segue 2                             & 149.4  & $-$38.1  & 35     & -- & -- \\  
  Ursa Major I                      & 159.4  & 54.4    & 97     & $18.3 \pm 0.24$ & \cite{Simon:2007dq}
  \footnotetext[1]{Galactic longitude and latitude.}
  \footnotetext[2]{\Jfactors are calculated assuming an NFW density
    profile and integrated over a circular region with a solid angle of $\Delta\Omega \sim 2.4
    \times 10^{-4} \sr$ (angular radius of 0.5\degree).}
  \footnotetext[3]{dSphs below the horizontal line are not included in the combined analysis.}
\end{tabular}
\end{ruledtabular}
\end{table}

\section{Acknowledgments}
The \textit{Fermi}-LAT Collaboration acknowledges support for LAT development, operation and data analysis from NASA and DOE (United States), CEA/Irfu and IN2P3/CNRS (France), ASI and INFN (Italy), MEXT, KEK, and JAXA (Japan), and the K.A.~Wallenberg Foundation, the Swedish Research Council and the National Space Board (Sweden). Science analysis support in the operations phase from INAF (Italy) and CNES (France) is also gratefully acknowledged.

\nocite{Ackermann:2012kca,Bregeon:2013qba,Portillo:2014ena,Mattox:1996zz,Rolke:2004mj,Jeffreys:1946,Burkert:1995yz,dePalma:2013pia,Carlson:2014nra,Strigari:2007at,Essig:2009jx}

\bibliographystyle{apsrev4-1}
\bibliography{bib}

\ifdefined\arxiv
\clearpage

\appendix

\setcounter{equation}{0}

\widetext
\begin{center}
  {\bf \large \large Supplemental Material: Searching for Dark Matter Annihilation
  from Milky Way Dwarf Spheroidal Galaxies with Six Years of Fermi-LAT
  Data}
\end{center}
\section{Pass 8 Event-Level Analysis}
\label{sec:pass8}

\passeight is a new event-level analysis for the LAT instrument and is
the successor to the \New{\passsevenrep} event-level analysis
\cite{Atwood:2013rka,Ackermann:2012kca,Bregeon:2013qba}.  Some of the
key features of \passeight are new algorithms to identify out-of-time
signals, a new tree-based pattern recognition for the tracker
subsystem, and an improved energy reconstruction that extends the LAT
energy range below 100\MeV and above 1\TeV.  \passeight implements a
new classification analysis based on boosted decision trees (BDTs),
which provides enhanced background rejection power relative to
\New{\passsevenrep}~\cite{Atwood:2013rka}.  The \passeight event-level analysis
enhances the capabilities of the LAT in all metrics relevant for
high-level science analysis.  In the energy range between 1\GeV and
10\GeV, the new P8R2\_SOURCE event class has a 30--40\% better
point-source sensitivity than the P7REP\_CLEAN event class.

\passeight introduces an \textit{event type} classification scheme
that partitions events within a class according to their
reconstruction quality.  The event type classification is a
generalization of the existing conversion type designation that
identifies events converting in the \textit{Front} or \textit{Back}
section of the tracker~\cite{Atwood:2009ez,Portillo:2014ena}.
\passeight defines eight new event type selections based on a sequence
of energy-dependent cuts on the BDT variables that categorize the
quality of the direction and energy reconstruction.  Four event types
categorize the quality of the directional reconstruction (PSF0 to
PSF3), and another four do so for the energy reconstruction (EDISP0 to
EDISP3).  By construction, these selections partition the gamma-ray
acceptance at each energy such that an event class will have
approximately the same number of events of each type.

Our maximum likelihood analysis of the dSphs combines the four
P8R2\_SOURCE\_V6 PSF event \New{types} in a joint likelihood
function.  Although each event type contains approximately the same
fraction of the total instrument acceptance, the angular resolution as
measured by the 68\% and 95\% containment radii of the PSF is significantly
better for events belonging to the best PSF event types.  At 3.16\GeV
the 68\% (95\%) containment radii of the \New{acceptance-weighted}
PSF for the best and worst PSF event types (PSF3 and PSF0) is
0.17\degree (0.35\degree) and 0.92\degree (2.3\degree).  By combining
the event types in a joint likelihood function, we weight the contribution of
events within a class by their reconstruction quality, e.g., events
with the least well-characterized direction (PSF0) are assigned the
lowest weight when testing the hypothesis of a putative DM source.
We estimate that splitting the event sample by event type improves the
sensitivity to an isolated point source by 10\%.  We expect that
larger sensitivity gains will occur in regions where the gamma-ray
intensity is strongly
non-uniform. 

Given these improvements, along with two additional years of data, our
flux constraints are expected to improve by a factor of $\sim$1.7 below
10\GeV and $\sim$2.2 above 100\GeV relative to the analysis of
\citet{Ackermann:2013yva}.  Although both the \New{\passsevenrep} and
\passeight analyses yield limits on the DM annihilation cross section
within their respective 95\% sensitivity bands, their constraints
differ by a factor that is appreciably larger than expected from the
median experimental sensitivities.  For the \bbbar channel, the
\passeight constraints are $\roughly 5$ times lower at $100\GeV$.  For
two independent data sets, statistical fluctuations can easily account
for the difference in limit realizations.

We find that, at a given energy, \New{only 20--40\%} of events in the
six-year \passeight SOURCE-class data set are shared with the
four-year \passsevenrep CLEAN-class data set.  \New{If the \passeight
  SOURCE-class selection retains all events in the \passsevenrep data
  set, we expect this fraction to equal the product of the ratio of
  gamma-ray acceptances for the two event classes
  with the ratio of observation times --- in this case 35--50\%.
  Since the basic event reconstruction in \passeight is fundamentally
  different, however, the characteristics of individual events change
  slightly.  Events near the threshold of any of our analysis cuts can
  migrate out of our data selection and be excluded from the
  \passeight analysis.  Specifically, 1--3\% of the \passsevenrep
  events were reconstructed outside of our ROIs, and 10--15\% were not
  deemed likely enough to be photons to be included in the \passeight
  SOURCE class.  \figref{overlap} shows the fraction of shared events
  as a function of energy for the set of dSphs in our combined
  analysis, along with a sample of events from the Earth limb,
  selected from time periods when the magnitude of the rocking angle
  of the LAT was greater than 52\degree.  The Earth's limb is an
  extremely pure photon source, and the fact that its shared fraction
  lies only slightly below the event class acceptance ratio indicates
  that the P8R2\_SOURCE selection has a high efficiency for retaining
  gamma-ray events in the P7REP\_CLEAN selection.  The lower shared
  fraction observed in the dSph ROIs can be attributed to
  misclassified charged-particle events that constitute \roughly
  30--40\% of the diffuse, high-latitude background between 1 and
  10~GeV.  These events typically lie near the boundary of the
  selections used to discriminate gamma rays from charged-particle
  backgrounds and are much more likely to migrate out of a given event
  class than a true gamma-ray event.
}



Using the fractional overlap, we can estimate the evolution of a
background fluctuation between analyses.  For example, the largest
\New{\passsevenrep} excess occurred for the \bbbar channel at masses between 10
and 25 GeV, and had a local significance of $\roughly 3\sigma$.  If we
assume this excess resulted from an upward fluctuation of the
background, the addition of new data is likely to reduce its
significance.  
\New{If the excess is due to a statistical fluctuation, we
  quantitatively expect the original significance, $\sigma_{1}$, to drop by a
  factor that depends only on the intrinsic fraction of shared events,
  $f_i$, and the ratio of observation times $t_{1}/t_{2}$,
\begin{equation}
  \sigma_{2} \approx \sigma_{1} f_{i} \sqrt{t_{1}/t_{2}},
\end{equation}
\noindent
where the intrinsic fraction is related to the observed fraction
($f_{obs}$) by $f_{i} \approx f_{obs}(t_{2}/t_{1})$.}  In the energy
range between 1 and 10\GeV, the \passeight analysis shares
\New{$f_{obs} \sim$35\%} of its events with \passsevenrep and has a 50\%
longer observation period corresponding to an intrinsic shared
fraction of \New{$f_{i} \sim$52\%}.  \New{A 3$\sigma$ background
  fluctuation should therefore drop to $\sim$1.3$\sigma$, which is
  consistent with our observations.}


\begin{figure}
\includegraphics[width=0.9\columnwidth]{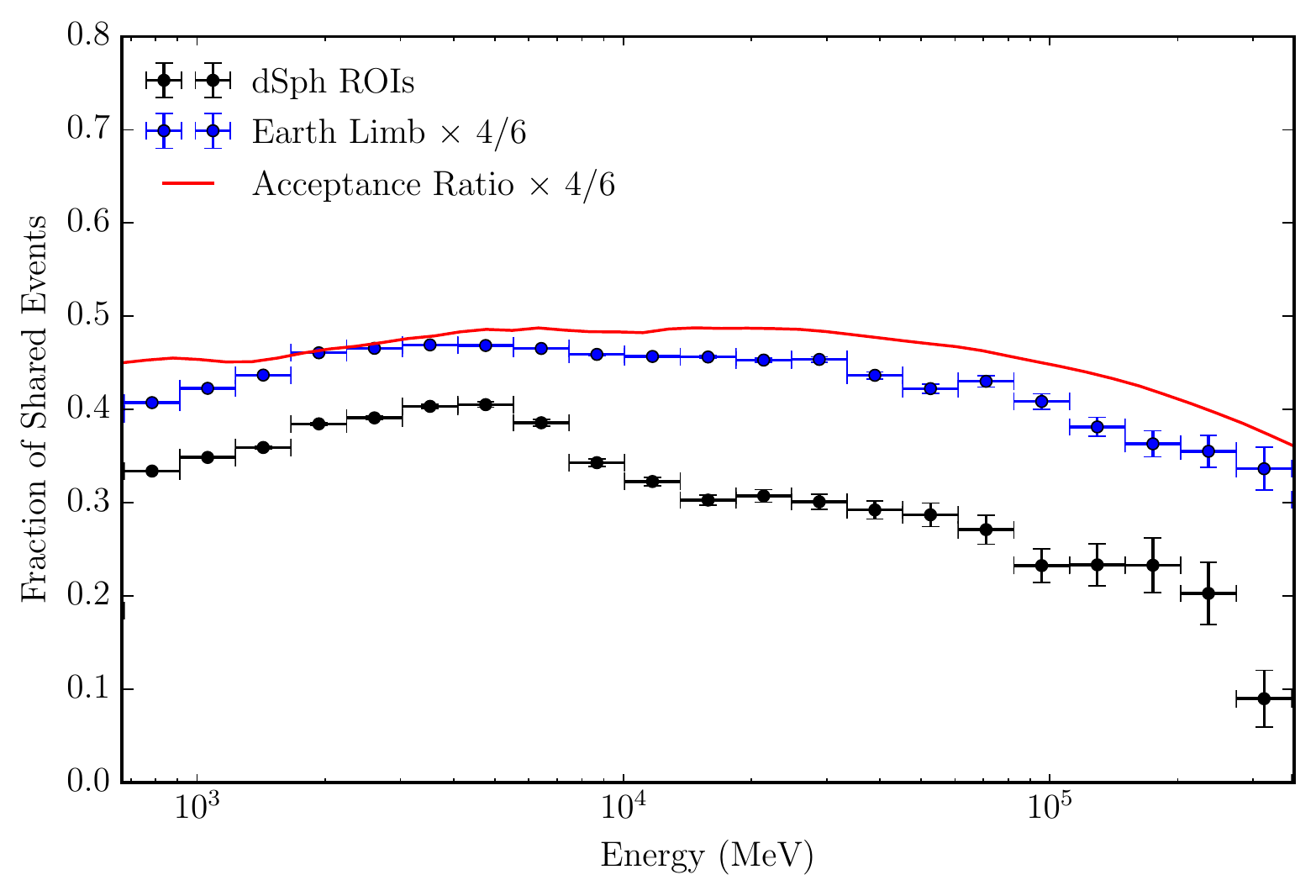}
\caption{\New{Fraction of events in the 6-year \passeight SOURCE data
    set that are also in the 4-year \passsevenrep CLEAN data set.  The
    lower points (black) show the shared fraction for events within
    $10\degree$ of one of the 15 dSphs used in the combined
    analysis.  
    The upper points (blue) show the expectation for the shared
    fraction of a pure gamma-ray sample as derived from events taken
    from the Earth Limb.  The red, solid line shows the acceptance
    ratio of P7REP\_CLEAN\_V15 to P8R2\_SOURCE\_V6 and represents the
    maximum possible shared event fraction.  Both the Earth Limb
    fraction and acceptance ratios have been scaled by the ratio of
    the observation times of the 4- and 6-year analyses.}
  \label{fig:overlap}}
\end{figure}

\New{
\section{Galactic Diffuse Background Model}
\label{sec:diffuse}

We model diffuse backgrounds in the dSph ROIs using templates that
account for both astrophysical backgrounds and residual particle
contamination.  Our model for the Galactic diffuse emission is based
on the \passsevenrep diffuse emission model
(\texttt{gll\_iem\_v05\_rev1.fit}) that was derived from a fit to
\passsevenrep data with the P7REP\_CLEAN\_V15 IRFs.  Although models
of the Galactic diffuse emission are formally independent of the IRFs,
the \passsevenrep model was fit without accounting for energy
dispersion which introduces some dependence on the \passsevenrep IRFs.
To create a model that can be used self-consistently with \passeight
data, we have derived a rescaled model (\texttt{gll\_iem\_v06.fit})
that accounts for the influence of energy dispersion by correcting for
the differences in apparent intensity in the \passsevenrep and
\passeight data sets.

The correction for energy dispersion is derived from the ratio of the
counts distributions computed with and without the correction for
energy dispersion.  Given an intensity, $I(E)$, the counts densities
evaluated with energy dispersion ($C_{1}$) and assuming perfect energy
resolution ($C_{2}$) are given by
\begin{equation}\label{eqn:edisp0}
\begin{aligned}
C_1(E') = \int\int I(E) \aeff(E,\theta) t_{\rm obs}(\theta) D(E';E,\theta)dEd\theta, 
\end{aligned}
\end{equation}
\begin{equation}\label{eqn:edisp1}
\begin{aligned}
C_2(E') = I(E) \int\int  \aeff(E,\theta) t_{\rm obs}(\theta) \delta(E' - E) dEd\theta,
\end{aligned}
\end{equation}
where $E$ and $E'$ are the true and measured energies, $\theta$ is the
LAT incident angle ($\theta=0$ is normal to the top of the LAT),
$\aeff(E,\theta)$ is the effective area, $t_{\rm obs}$ is the
integrated livetime, and $D(E';E,\theta)$ is the energy dispersion.
Defining ratios between the counts densities as $R_{P8} = C_1^{P8} /
C_2^{P8}$ and $R_{P7REP} = C_1^{P7REP} / C_2^{P7REP}$, the rescaled
model is $I_{P8}(E) = (R_{P8} / R_{P7REP})I_{P7REP}(E)$.  We note that
the correction depends on the true intensity ($I(E)$) in Equations
\ref{eqn:edisp0} and \ref{eqn:edisp1}.  We find that the correction is
not strongly dependent on the assumed spectrum, and we use here the
average all-sky intensity of the \passsevenrep model.  Systematic
uncertainties associated with our model for the Galactic diffuse
emission are discussed in more detail in the Systematic Uncertainties
section.}


\section{Statistical Methodology}
\label{sec:stats}

We use a maximum likelihood-based statistical formalism
\cite{Mattox:1996zz} to test the DM signal hypothesis and derive
confidence intervals on \sigmav.  Our global likelihood function for
\sigmav is constructed from the product of likelihood functions for
individual dSphs in our sample.
We compute the profile likelihood function \New{as a function of}
\sigmav by maximizing the global likelihood function with respect to
the nuisance parameters for each dSph ($\nuisance_{i} =
\{\params_{i},J_{i}\}$):
\New{
\begin{equation}\label{eqn:global_likelihood}
\lambda(\sigmav,\mDM) = 
\prod_{i}\pseudolike_i(\sigmav,\mDM,\hat{\params}_{i}, \doublehat{J_{i}}(\sigmav,\mDM)\given\data_{i}).
\end{equation}
Here, $\hat{\params}_{i}$ are the best-fit parameters derived from a
global fit with a free-normalization DM source with a $\Gamma=2$
power-law spectrum and $\doublehat{J_{i}}$ is the \Jfactor that
maximizes $\pseudolike_i$ for a given \sigmav and \mDM.}
Confidence intervals on \sigmav \New{ for a given \mDM} are calculated
with the delta-log-likelihood technique, requiring a change in the
profile log-likelihood of 2.71/2 from its maximum for a 95\% \CL upper
limit \cite{Rolke:2004mj}.

Among the nuisance parameters in \eqnref{global_likelihood}, we
distinguish between parameters constrained by the gamma-ray data
(\params) and the \Jfactors ($J$), which are constrained by an
independent analysis of stellar kinematics.
We use \Jfactors derived from a two-level Bayesian hierarchical
modeling analysis that incorporates information on both stellar
kinematics and priors on the distribution of global dSph properties
\cite{Martinez:2013ioa}.  For each dSph the Bayesian analysis provides
a posterior distribution function, \ProbJi, which we approximate with
a log-normal distribution with central value, \Jobsi, and uncertainty,
\Jsigma.

Following the approach developed in
\citet{Ackermann:2011wa,Ackermann:2013yva}, we account for statistical
uncertainty on the \Jfactor by multiplying the LAT likelihood function
with a \Jfactor likelihood function,
$\Jlike(\Jtrue\given\Jobsi,\Jsigma)$.  We construct an \textit{ansatz}
for the \Jfactor likelihood function by equating the sampling
distribution of \Jobsi with \ProbJi.  With this underlying assumption,
the likelihood function is given by a log-normal distribution with central
value \Jobsi and width \Jsigma,
\begin{equation}\label{eqn:jfactor_likelihood2}
  \Jlike(\Jtrue \given \Jobsi, \Jsigma) = \frac{1}{\ln(10) \Jobsi
    \sqrt{2 \pi} \sigma_i} 
  e^{-(\logtenJtrue-\logtenJobs)^2/2\sigma_i^2}.
\end{equation}
We note that \citet{Ackermann:2011wa,Ackermann:2013yva} used a
different form for the \Jfactor likelihood function with
$\Jlike(\Jtrue\given\Jobsi,\Jsigma) = \ProbJi$ --- \ie, a log-normal
posterior.  The \Jfactor likelihood function used in this work differs
in the substitution of nominal \Jfactor (\Jobsi) for the true \Jfactor
(\Jtrue) in the denominator of \eqnref{jfactor_likelihood2}.  The
log-normal likelihood formulation has several advantages over the
log-normal posterior used in
\citet{Ackermann:2011wa,Ackermann:2013yva}.  When interpreted as a
sampling distribution for \Jobsi, it is properly normalized for all
values of \Jtrue.  The maximum likelihood estimator $\hat{\Jtrue}$ for
the \Jfactor also coincides with its nominal value \Jobsi from the
stellar kinematic analysis.

To confirm that our upper limits have the correct frequentist
statistical coverage we have performed a series of independent Monte
Carlo realizations of our analysis in which we include a DM
signal.  In these simulations the true \Jfactors are fixed to their
nominal values while the measured \Jfactors are randomized by sampling
from a log-normal approximation to the \Jfactor posterior of each dSph.
\figref{coverage_scan} shows the upper limits on \sigmav from one set
of realizations simulated with a \bbbar annihilation spectrum and \mDM
= 25\GeV.  Under the assumption that the \Jfactor posterior is a good
representation of the sampling distribution for \Jfactor measurements,
we find that our statistical methodology produces the correct
statistical coverage for a 95\% \CL upper limit.

\begin{figure}
\includegraphics[width=0.9\columnwidth]{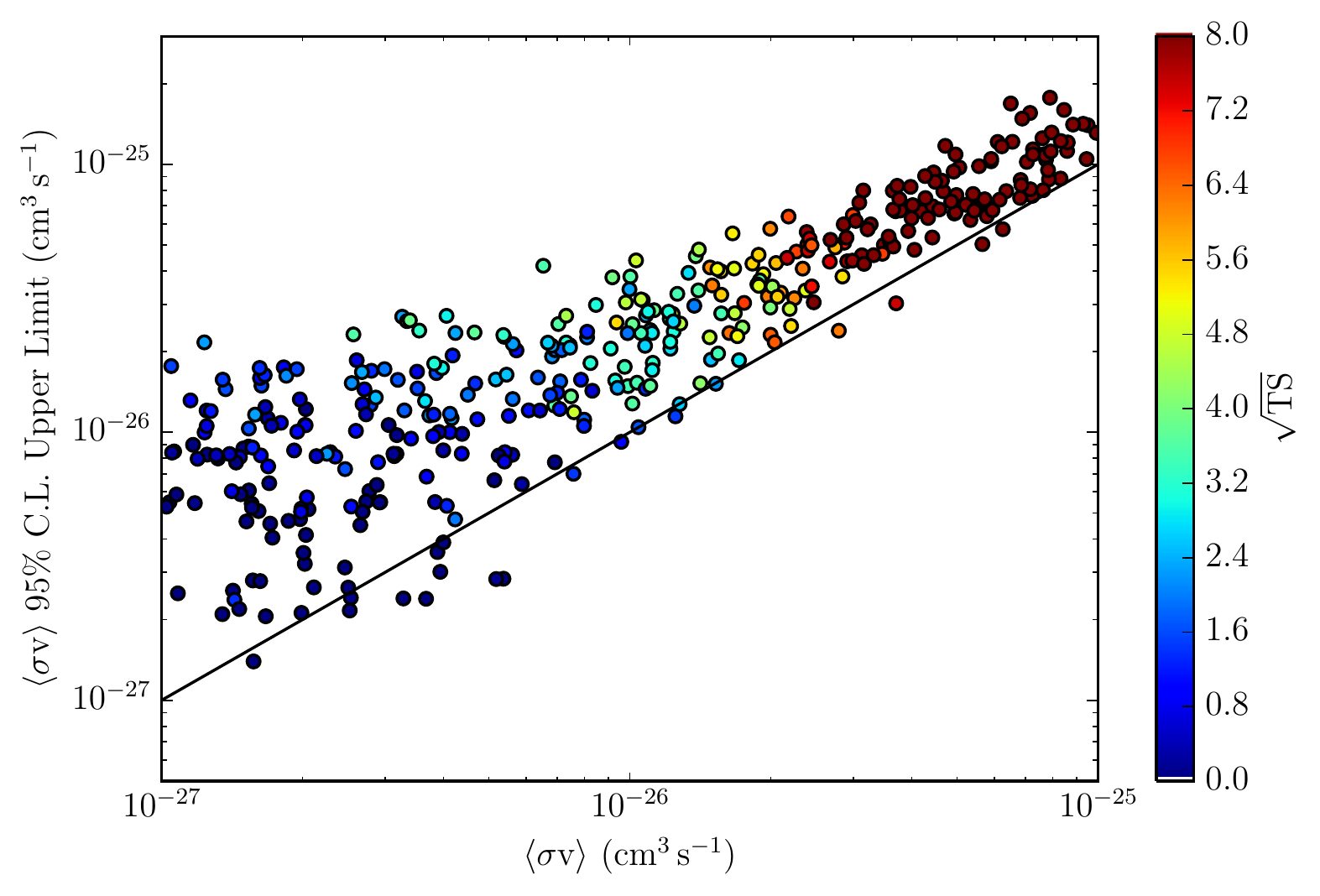}
\caption{Comparison of the 95\% \CL upper limit on \sigmav with its
  true value for a set of 400 MC realizations of the combined
  dSph analysis.  In each realization a DM model with \mDM = 25~GeV
  and a \bbbar annihilation spectrum is injected at the locations of
  the dSphs using the set of nominal dSph \Jfactors. 
  The DM cross-section is uniformly randomized in $\log\sigmav$ between
  $10^{-27}$ and $10^{-25}~\cmcubes$.  The measured \Jfactor (\Jobsi)
  of each dSph is randomized by sampling from the \Jfactor posterior.
  The color of each point indicates the square root of the DM test
  statistic evaluated with the same spectrum as the injected DM
  model. Points above the solid diagonal line represent realizations
  where the upper limit covers the true value of the injected
  signal.}
\label{fig:coverage_scan}  
\end{figure}

\section{Hybrid Bayesian Analysis}

The main results of this work are evaluated with the
delta-log-likelihood method, a fully frequentist statistical approach.
In constructing the likelihood function of the delta-log-likelihood
analysis, a central assumption is that the posterior distribution
function is a good approximation to the \Jfactor sampling
distribution.  If this assumption holds, then our limits have the
correct frequentist statistical coverage.  To determine the robustness
of our results to the choice of statistical methodology, we have
performed an alternative analysis based on a Bayesian statistical
approach in which we marginalize over the posterior distributions of
the \Jfactors.  For each target we use the same LAT likelihood
function as for the primary analysis but set $\Jlike = \ProbJi$.  
\New{We then marginalize over the \Jfactors to derive a
  one-dimensional marginal posterior density,}
\begin{equation}\label{eqn:sigmav_posterior}
  \Prob(\sigmav) = \frac{\int \prod_{i}\lambda_i(\sigmav,\mDM,J_i)\pi(\sigmav)dJ_{i}}
  {\int\prod_{i}\lambda_i(\sigmav',\mDM,J_{i}')\pi(\sigmav')d\sigmav' dJ_{i}'},
\end{equation}
where $\pi(\sigmav)$ is the prior for \sigmav and
\New{
\begin{equation}\label{eqn:sigmav_plike}
  \lambda_i(\sigmav,\mDM,J_i) = 
  \like_i(\sigmav,\mDM,J_i,\hat{\params}_i)\times \Prob(\Jtrue)
\end{equation}
is the product of the likelihood for target $i$ with its
\Jfactor posterior.  }
Given the marginal posterior density of \eqnref{sigmav_posterior}, we
derive an upper limit by finding the value $\sigmav_{0}$ that
satisfies $\int_{\sigmav_{0}}^{\infty}\Prob(\sigmav)d\sigmav = p$
where we use $p = 0.05$ to define a Bayesian equivalent to the
frequentist 95\% \CL upper limit.

An important consideration for the Bayesian analysis is the choice of
the prior distribution, $\pi(\sigmav)$, which is needed to evaluate
the posterior density in \eqnref{sigmav_posterior}.  In order to
choose a prior that minimally influences our inference on \sigmav, we
consider the class of non-informative priors derived according to
Jeffreys' rule~\cite{Jeffreys:1946}.  As two approximations to the
Jeffreys' prior for our likelihood, we take the Jeffreys' prior for
the mean of a Gaussian distribution of known width, the uniform prior
with $\pi(\mu)=1$, and the Jeffreys’ prior for a Poisson distribution,
$\pi(\mu)=\mu^{-1/2}$, which we refer to here as the Poisson prior.
The uniform prior should be applicable when the expected background is
large relative to the signal and the LAT sensitivity is
background-limited.  In this regime the likelihood function of the LAT
data given the model asymptotically approaches a Gaussian
distribution.  On the other hand, the Poisson prior is applicable when
the expected background is negligible and the likelihood is well
approximated by a Poisson distribution.  For spectral models of WIMP
annihilation through quark or lepton channels, the background- and
signal-limited sensitivity regimes correspond to models of low and
high mass, respectively.

\figref{bayesian_limits} compares limits for the \bbbar channel
calculated with the delta-log-likelihood and Bayesian analyses.  In
this comparison we calculate two sets of Bayesian upper limits using
the \sigmav marginal posterior (Equation~\ref{eqn:sigmav_posterior})
and substituting the uniform and Poisson priors for $\pi(\sigmav)$.
We find that the Bayesian upper limits are in good agreement with the
limits of the delta-log-likelihood analysis when the appropriate prior
is chosen for the form of the likelihood on \sigmav.  For DM masses
below 100~GeV where the likelihood is well approximated by a Gaussian,
the limits from the Bayesian analysis with a uniform prior lie within
10\% of those from the delta-log-likelihood analysis.  At higher DM
masses, a similar level of agreement (10--20\%) is observed when
comparing the delta-log-likelihood limits to the limits evaluated with
the Poisson prior.  We note that these changes are comparable to or
smaller than the effect of the systematic uncertainties considered in
the following sections.  We conclude that our upper limits are robust
to the choice of statistical methodology used to model the \Jfactor
uncertainties.

\begin{figure}
\includegraphics[width=0.9\columnwidth]{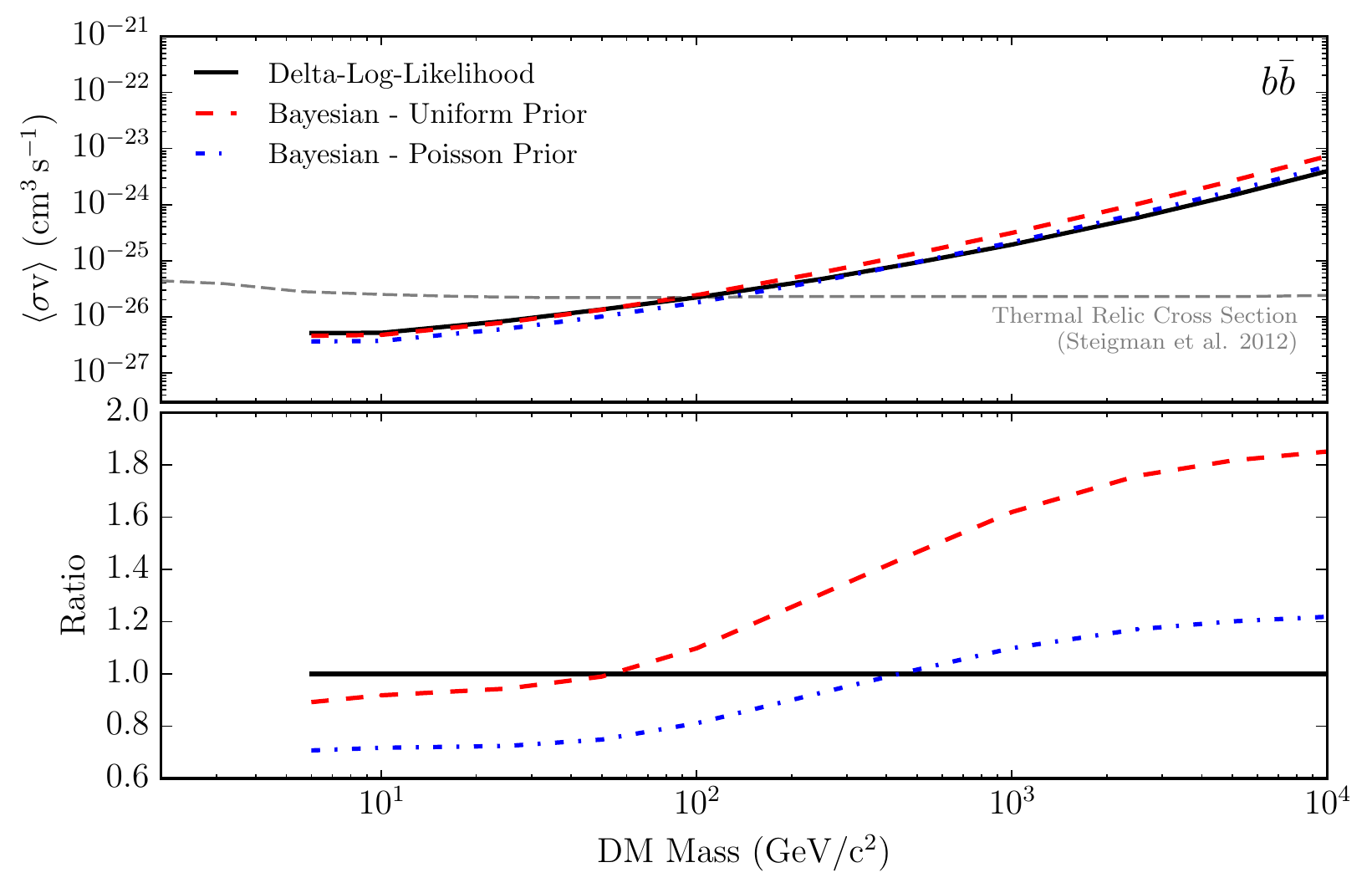}
\caption{Comparison of upper limits (\bbbar channel) for the combined
  analysis of 15 dSphs as derived with the delta-log-likelihood
  analysis (solid line) and the Bayesian analysis performed with a
  uniform (dashed line) and Poisson (dot-dashed line) prior.  The
  lower panel shows the ratio of these curves to the limits for the
  delta-log-likelihood analysis.}
\label{fig:bayesian_limits}
\end{figure}

\section{Systematic Uncertainties}
The dominant systematic uncertainties of this analysis arise from incomplete knowledge in three areas: the LAT instrument response, the Galactic diffuse gamma-ray background, and the distribution of DM in the dSphs. 
To estimate the impact of these uncertainties, we repeat our DM search using varying assumptions intended to encompass the range of possibilities in each of these areas.
Below we address systematics associated with the IRFs and diffuse background model, which both affect constraints at the 10\% level, with the latter becoming less relevant for hard DM spectra ($\mDM >100 \GeV$).
Systematics associated with the \Jfactors are addressed in the following section, while here we quote the maximum deviation from our fiducial NFW model, which occurs when assuming a cored Burkert density profile \cite{Burkert:1995yz},
\begin{equation}
 \label{eq:burkert}
 \rho_{\DM}(r) = \frac{\rho_0 r_s^3}{ (r_s + r) ( r_s^2 + r^2) }.
\end{equation}
The \Jfactor systematic uncertainty has a greater impact than that of
the IRF or diffuse models, approximately 35\% at 100 GeV.  We provide
a summary of the systematic uncertainty as a function of DM mass and
annihilation channel in \tabref{systematics}.

In addition to the standard model of interstellar gamma-ray emission
for the LAT, we consider eight alternative models to sample a fairly
wide range of possibilities for the diffuse gamma-ray background
\cite{dePalma:2013pia}.  Although we can vary parameters \emph{within}
our background models, there are no doubt sources of gamma-ray
emission that remain unmodeled.  It was observed by
\citet{Ackermann:2013yva} that the TS distribution from random blank
sky locations deviated from statistical expectations, suggesting an
incomplete background model.
This indicated that a rescaling was necessary when converting from TS
to significance, effectively lowering the sensitivity of the study.
One large class of objects known to be unmodeled are sub-threshold
point sources, \ie, those which contribute gamma rays but are not
significant enough individually to be included in a catalog.  It has
been speculated that these give rise to the larger than expected rate
of type I errors (false positives) that skew the TS distribution
relative to the expectation from Poisson statistics
\cite{Ackermann:2013yva,Carlson:2014nra}.

\figref{individual_ts} shows the distribution of TS obtained from the
analysis of randomized ROIs when the data are analyzed using the 2FGL
and 3FGL catalogs.  We additionally analyze simulated ROIs with the
3FGL catalog using an input model for the simulations that includes
3FGL sources and our templates for the Galactic and isotropic diffuse
backgrounds.  Using the 3FGL, which roughly doubles the number of
modeled sources, brings the TS distribution closer to the asymptotic
expectation.  However, a significant deviation with respect to the
asymptotic expectation from Chernoff's theorem is still observed,
indicating that additional unmodeled components may still be present
in the data.

The uncertainty in the LAT response is bracketed by using IRFs that are maximally and minimally sensitive to our signal.  The maximally sensitive set has a greater effective area, narrower PSF, and accounts for dispersion in the reconstructed energy.  The minimally sensitive IRFs are the opposite.  The effective area is set at the boundaries of the envelope described in \citet{Ackermann:2012kca}, while the energy dispersion and PSF width are scaled by $\pm 5\%$ and $\pm 15\%$, respectively.

\begin{table}[h] \scriptsize
  \caption{\label{tab:systematics}  Effect of systematic uncertainties
    for various WIMP masses and channels reported as a symmetrical 
    relative deviation from the combined 95\% \CL upper limits.}
\begin{ruledtabular}
\begin{tabular}{ l c c c c c }
 &  & 10\GeV  & 100\GeV  & 1\TeV  & 10\TeV \\
\hline
\hline 
\multirow{3}{*}{$e^{+}e^{-}$}  & IRFs & 6\% & 10\% & 11\% & 11\% \\ 
 & Diffuse & 12\% &  6\% &  3\% &  2\% \\ 
 & \Jfactor & 29\% & 31\% & 17\% & 16\% \\
\hline 
\multirow{3}{*}{$\mu^{+}\mu^{-}$}  & IRFs & 6\% & 10\% & 11\% & 11\% \\ 
 & Diffuse & 13\% &  6\% &  3\% &  2\% \\ 
 & \Jfactor & 28\% & 32\% & 18\% & 16\% \\
\hline 
\multirow{3}{*}{$\tau^{+}\tau^{-}$}  & IRFs & 7\% & 9\% & 11\% & 11\%\\ 
 & Diffuse & 15\% &  6\% &  1\% & 1\% \\ 
 & \Jfactor & 24\% & 35\% & 15\% & 14\% \\
\hline 
\multirow{3}{*}{$u \bar u$}  & IRFs & 6\% & 7\% & 9\% & 10\%\\ 
 & Diffuse & 23\% & 12\% &  7\% &  4\% \\ 
 & \Jfactor & 16\% & 34\% & 31\% & 24\% \\ 
\hline 
\multirow{3}{*}{$b \bar b$}  & IRFs & 6\% & 7\% & 9\% & 11\% \\
 & Diffuse & 23\% &  13\% &  7\% &  4\% \\
 & \Jfactor & 13\% & 32\% & 32\% & 23\% \\
\hline 
\multirow{3}{*}{$W^{+}W^{-}$}  & IRFs & & 7\% & 10\% & 11\%\\ 
 & Diffuse &  & 13\% & 6\% & 2\% \\ 
 & \Jfactor & & 32\% & 31\% & 17\% \\ 

\end{tabular}
\end{ruledtabular}
\end{table}

\begin{figure}
\includegraphics[width=0.49\columnwidth]{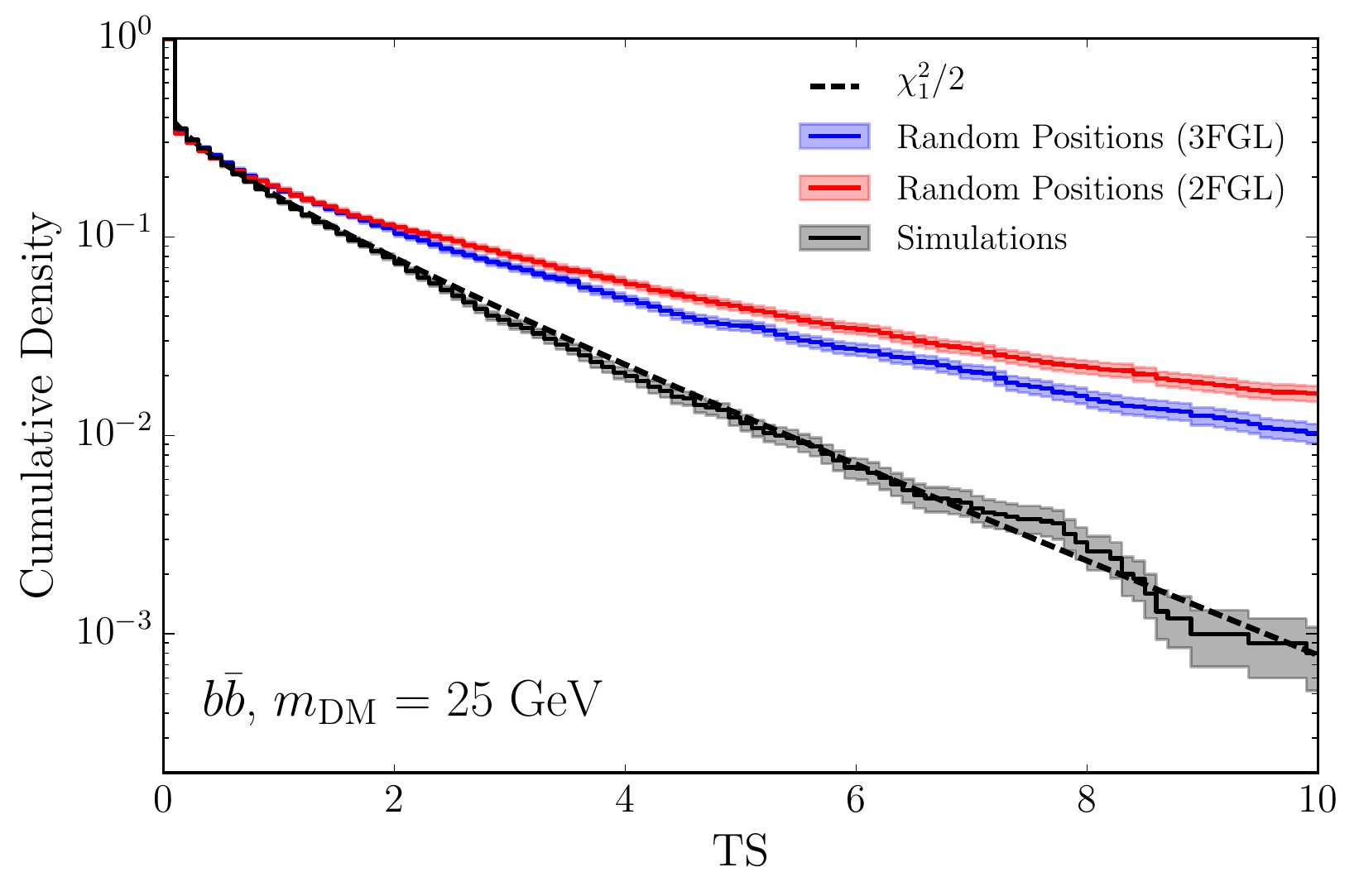}
\includegraphics[width=0.49\columnwidth]{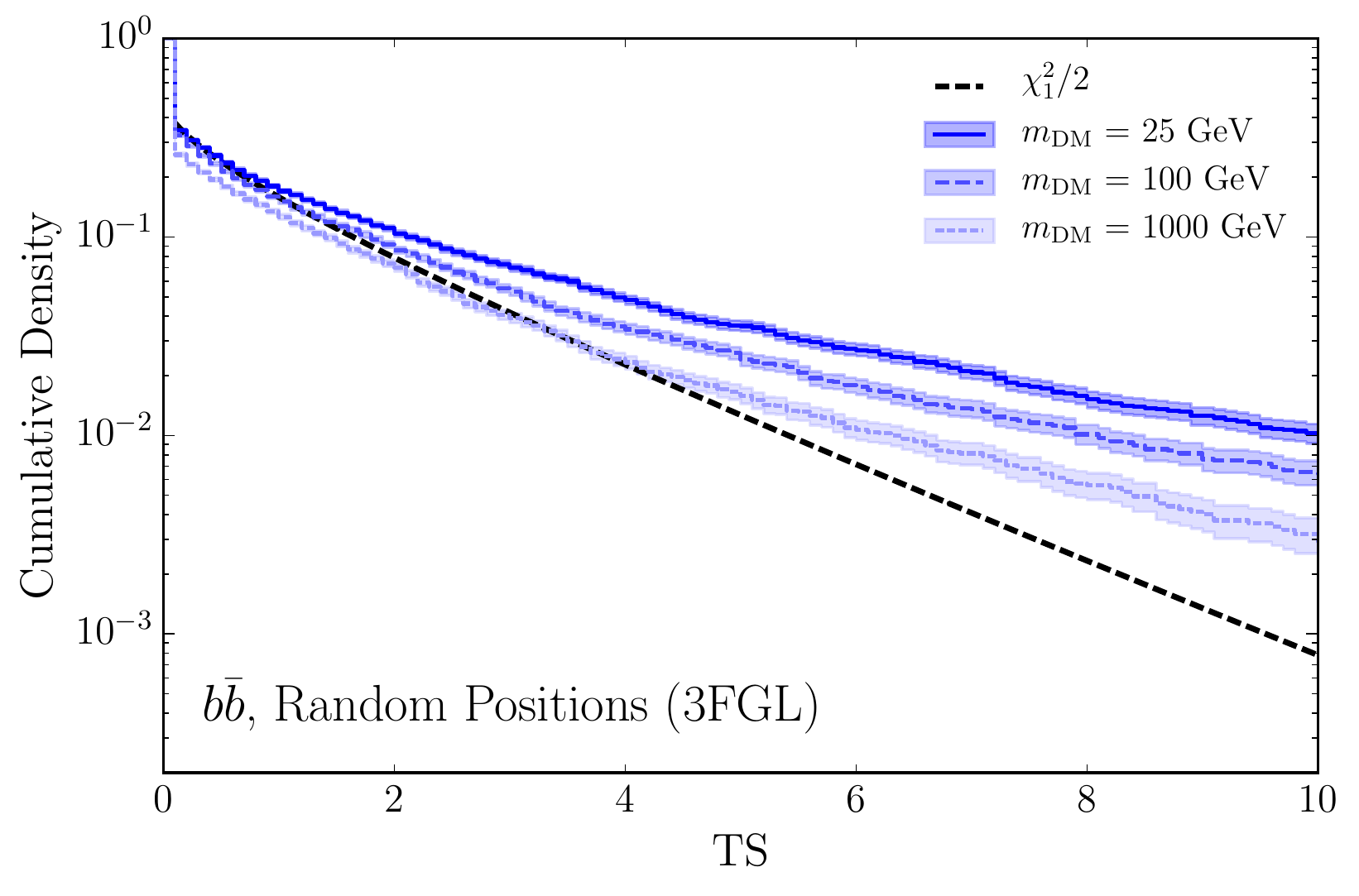}
\caption{\New{Cumulative distribution of TS from 7500 randomized
    blank-sky regions fit with a DM particle annihilating through the
    \bbbar channel.  Shaded bands indicate the one sigma uncertainties
    on the cumulative fraction, which are highly correlated between
    bins. \textit{Left panel:} Distributions for a DM mass of 25\GeV
    evaluated from Monte Carlo simulations (black line) and LAT data
    analyzed with the 3FGL (blue line) and 2FGL (red line)
    point-source catalogs. \textit{Right panel:} Distributions for DM
    masses of 25\GeV, 100\GeV, and 1000\GeV evaluated with LAT data
    analyzed with the 3FGL point-source catalog.  As illustrated by
    this plot, the TS distribution depends on the parameters of the
    WIMP spectral model (annihilation channel and mass).  Harder
    spectral models (\eg, those with higher mass) have a distribution
    that lies closer to the asymptotic expectation from Chernoff's
    theorem.}}\label{fig:individual_ts}
\end{figure}


\section{J-factor Uncertainties}
\label{sec:jfactors}

We have described a statistical methodology to account for
uncertainties on the \Jfactors when deriving limits on the DM
annihilation cross section.
This procedure captures statistical uncertainties on the \Jfactor
arising from the analysis of stellar velocity dispersion data.  As
implemented in the likelihood, it is not intended to account for
additional systematic uncertainties.  Such systematic uncertainties on
the \Jfactor include parameterization of the DM profile and the choice
of priors for the profile parameters.  While previous studies have
shown that the derived \Jfactors are robust against these systematic
uncertainties for dSphs with large stellar data
sets~\cite{Strigari:2007at}, it is nonetheless important to quantify
their impact.  To assess the impact of systematic uncertainties in the
\Jfactor derivations, we examine a set of four alternative \Jfactors
derived by various fitting methods.

The first set of alternative \Jfactors comes from the recent analysis
of \citet{Geringer-Sameth:2014yza} assuming a generalized NFW profile
with non-informative priors on its parameters.  We also examine the
\Jfactors derived by \citet{Charbonnier:2011ft} assuming a generalized
Hernquist profile with uniform priors.
Additionally, we perform our own alternative analysis following the
procedure of \citet{Essig:2009jx} assuming a simple NFW profile with
non-informative priors on the scale radius and scale density, and a
velocity anisotropy parameter that is assumed to be constant with
radius.  Lastly, we show results derived from the multi-level modeling
approach of \citet{Martinez:2013ioa} assuming a cored Burkert profile
as presented in \citet{Ackermann:2013yva}.

For each of these alternative sets of \Jfactors, we re-derive the
limit on the DM annihilation cross section in the context of the LAT
data.  For cuspy spatial profiles, the spatial template of the DM
distribution has little impact on the LAT analysis.  Thus, for the
first three sets of alternative \Jfactors we only alter the nominal
\Jfactor and associated uncertainty.  When assuming a Burkert profile,
we use the full spatial profile of the assumed DM distribution (the
change in spatial profile affects the limits by ${<}\,5\%$).  Since
the analyses of \citet{Charbonnier:2011ft} and \citet{Geringer-Sameth:2014yza}
do not include all of the dwarfs used in our analysis, when a dSph
is missing from one of these data sets we assign it the nominal
\Jfactor and uncertainty from \tabref{dsphs} in the main text.  When
asymmetric errors are given for the best-fit \Jfactor, we use the
geometric mean to set the width of the log-normal \Jfactor likelihood.

The resulting change in the upper limit on the DM cross
section is shown in \figref{jfactor_syst}.  The mass dependence of the
curves in \figref{jfactor_syst} reflects the fact that by changing the
\Jfactors we change the relative importance of each dSph,
leading to an interplay between the LAT data and the assumed
\Jfactors.  Unsurprisingly, the largest change in the upper limit
comes from requiring a cored Burkert profile.  This increases the
upper limit by a factor of 20--40\% with respect to the nominal limit
(this is slightly larger than was observed by
\citet{Ackermann:2013yva}) and is what we quote in \tabref{systematics} as the overall \Jfactor systematic uncertainty.
The \Jfactors derived by \citet{Charbonnier:2011ft} and the alternative analysis with non-informative priors both yield slightly smaller changes in the
limit.  Finally, we observe that the \Jfactors from
\citet{Geringer-Sameth:2014yza} are most similar to the nominal \Jfactors and result in differences of 5--10\%.

The combined limits presented here include both classical and
ultra-faint dSphs.  Bayesian hierarchical modeling sets rather tight
constraints on the \Jfactors of the ultra-faint dSphs as members of
the dSph population; however, stellar kinematic data yield larger
uncertainties on ultra-faint dSphs when analyzed individually.  To
assess the maximum impact of mis-modeling the ultra-faint dSphs, we
split the dSph population into ultra-faint (Bootes~I, Canes
Venatici~II, Coma Berenices, Hercules, Leo~IV, Segue~1, Ursa Major~II,
Willman~1) and classical (Carina, Draco, Fornax, Leo~II, Sculptor,
Sextans, Ursa Minor) galaxies.  For soft annihilation spectra (e.g.,
the \bbbar channel for DM with mass ${<}\,100\GeV$), the classical and
ultra-faint populations yield comparable limits, each $\roughly 40\%$
worse than the combined limit.  
For harder annihilation spectra with spectral energy distributions
that peak above 10\GeV, the limits from the ultra-faint population are
roughly comparable to the combined limits, while the classical dSphs
yield limits up to five times weaker.  Considering only the classical
dSphs, models with the thermal relic cross section are excluded for
slightly lower masses (${\lesssim}\,80\GeV$).

\begin{figure}
\includegraphics[width=0.9\columnwidth]{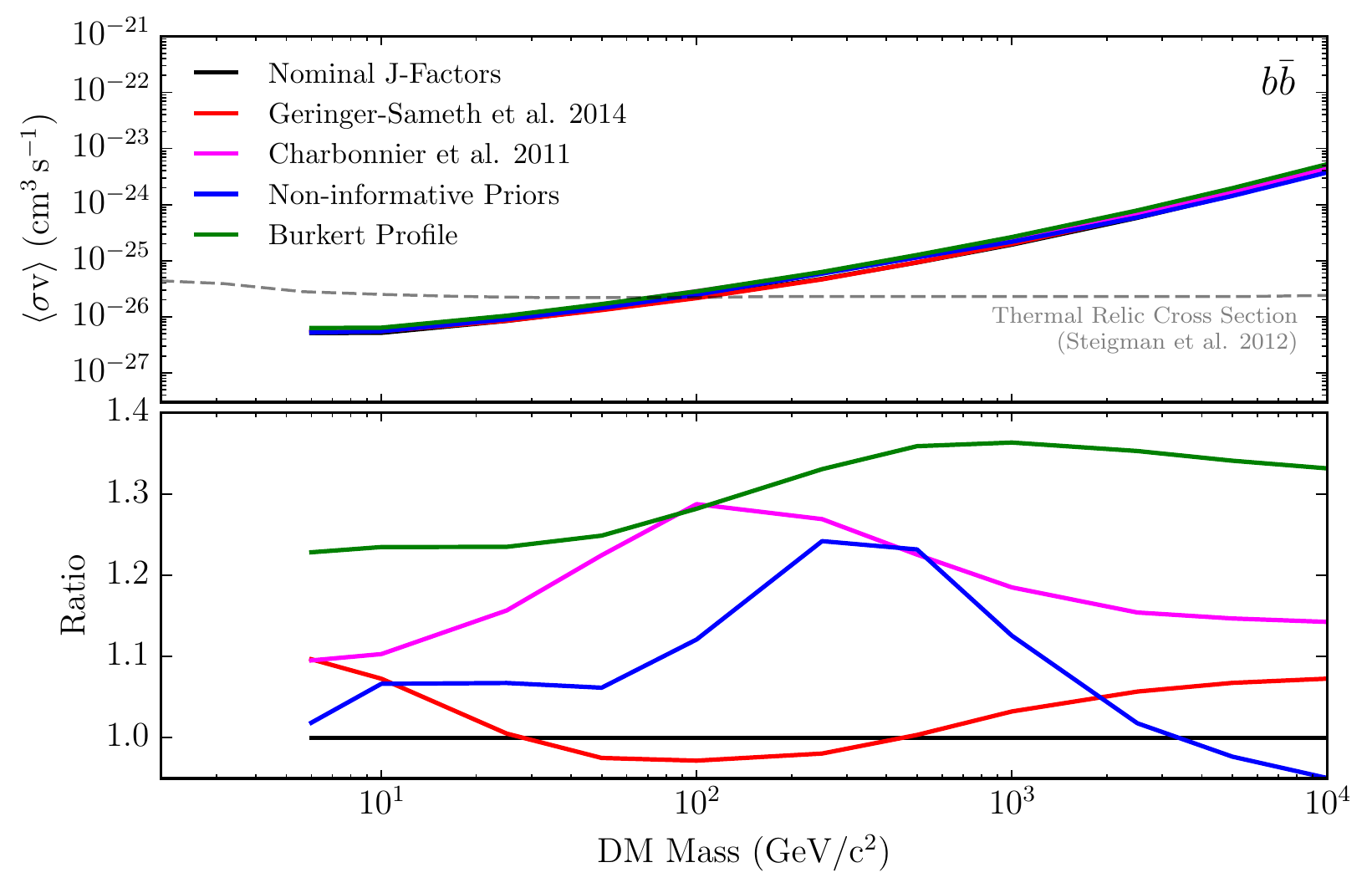}
\caption{Change in the limits derived for the DM annihilation cross
  section under the assumption of alternative sets of
  \Jfactors. Alternative \Jfactors are taken from
  \citet{Geringer-Sameth:2014yza} and
  \citet{Charbonnier:2011ft}. Non-informative priors are used to
  derive \Jfactors following the procedure of
  \citet{Essig:2009jx}. Burkert \Jfactors are derived using the
  multi-level modeling approach of \citet{Martinez:2013ioa} and are
  taken from \citet{Ackermann:2013yva}. }\label{fig:jfactor_syst}
\end{figure}

\section{Annihilation Channels}
\label{sec:channels}
WIMPs may annihilate through a variety of Standard Model channels.
For the quark and boson channels, the resulting gamma-ray spectra are
all similar and largely depend on \mDM.  The three leptonic channels have
harder spectral energy distributions with a peak in energy flux that is
closer to \mDM.  We perform our analysis for six representative
annihilation channels (\bbbar, \tautau, \mumu, \ee, \ww, and \uubar)
and for each we assume a 100\% branching fraction.  The resulting
constraints, shown in \figref{all_limits}, are similar to the
\bbbar and \tautau channels depicted in the main body of this work,
except for the \ee and \mumu channels which are somewhat higher.

\begin{figure}
  \includegraphics[width=0.9\columnwidth]{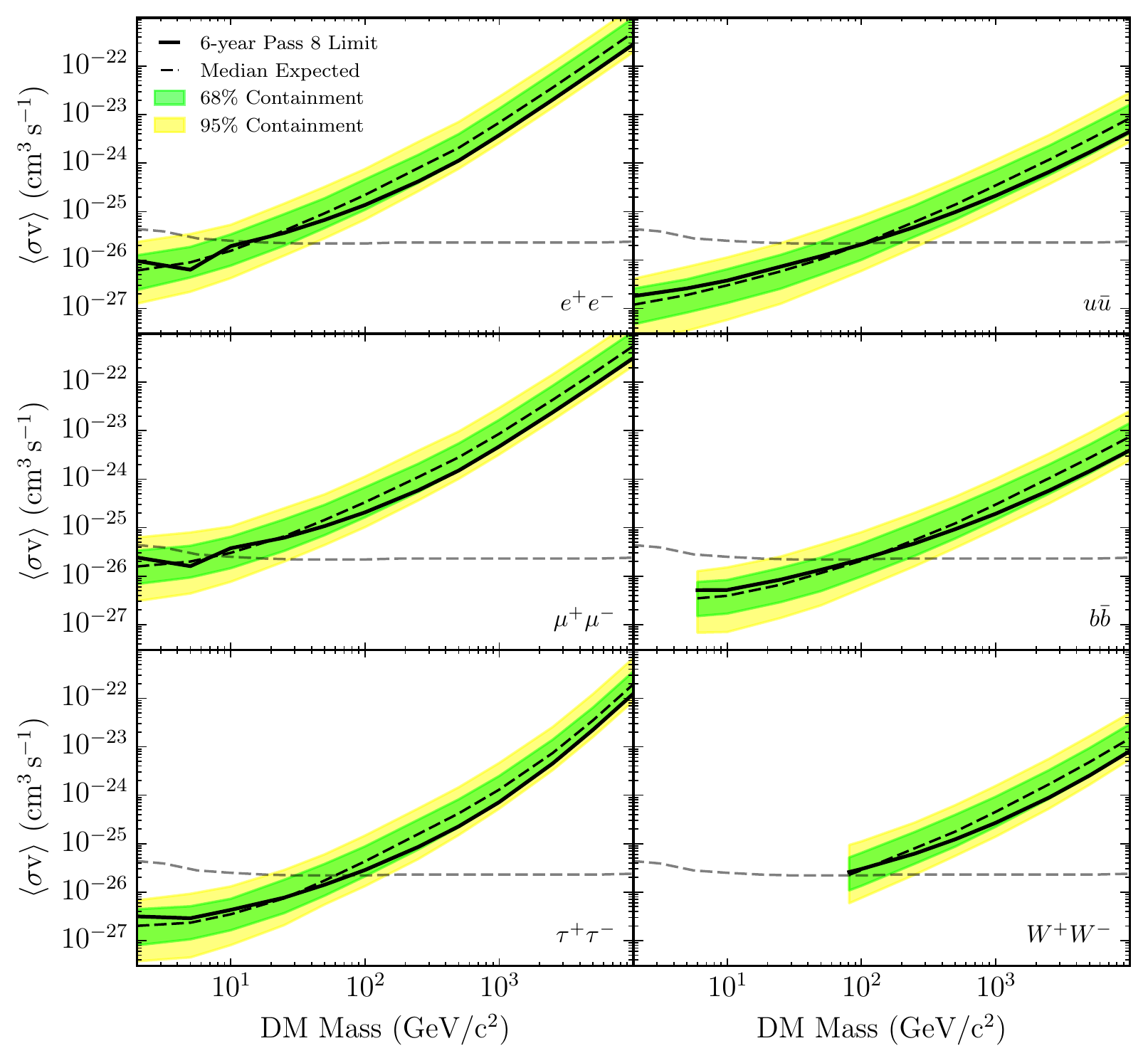}
  \caption{DM annihilation cross-section constraints derived from the
    combined 15-dSph analysis for various
    channels.}\label{fig:all_limits}
\end{figure}

\fi


\end{document}